\documentclass[useAMS,usenatbib]{mnras}

\usepackage{amssymb}
\usepackage{amsmath}
\usepackage{amsfonts}
\usepackage{graphicx}
\usepackage{subfig}
\usepackage{color}
\usepackage{multirow}

\def\aj{AJ}

\def\apj{ApJ}
\def\apjl{ApJ}
\def\apjs{ApJS}
\def\aap{A\&A}
\def\mnras{MNRAS}
\def\nat{Nature}
\def\pasp{PASP} 

\def\aaps{A\&AS}

\def\FeH{\mathrm{[Fe/H]}}
\def\aFe{[\alpha/\mathrm{Fe}]}
\def\e{{\rm e}}
{\newif\ifnotend
\notendtrue
\def\veclist{ABCDEFGHIJKLMNOPQRSTUVWXYZabcdefghijklmnopqrstuvwxyz.}
\def\top#1#2.{#1}
\def\tail#1#2.{#2.}
\loop\expandafter\xdef\csname v\expandafter\top\veclist\endcsname%
{{\noexpand\bf\expandafter\top\veclist}}
\edef\veclist{\expandafter\tail\veclist}
\if\veclist.\notendfalse\fi\ifnotend\repeat}

\def\kpc{\,{\rm kpc}}

\def\kms{\,\mathrm{km\,s}^{-1}}

\newcommand{\degree}{\ensuremath{^\circ}}
\newcommand{\diff}{\mathrm{d}}
\def\FeH{\mathrm{[Fe/H]}}
\def\aFe{[\alpha/\mathrm{Fe}]}
\def\e{{\rm e}}
{\newif\ifnotend
\notendtrue
\def\veclist{ABCDEFGHIJKLMNOPQRSTUVWXYZabcdefghijklmnopqrstuvwxyz.}
\def\top#1#2.{#1}
\def\tail#1#2.{#2.}
\loop\expandafter\xdef\csname v\expandafter\top\veclist\endcsname%
{{\noexpand\bf\expandafter\top\veclist}}
\edef\veclist{\expandafter\tail\veclist}
\if\veclist.\notendfalse\fi\ifnotend\repeat}

\def\kpc{\,{\rm kpc}}

\def\kms{\,\mathrm{km\,s}^{-1}}

\title[The age gradient in halo BHBs]{Characterizing stellar halo populations II: The age gradient in blue horizontal-branch stars}

\author[Payel Das]{Payel Das $^{1}$\thanks{E-mail:payel.das@physics.ox.ac.uk}, {Angus Williams $^{2}$}, and {James Binney $^{1}$}\\
$^{1}$Rudolf Peierls Centre for Theoretical Physics, University of Oxford, OX1 3NP, UK\\
$^{2}$Institute of Astronomy, University of Cambridge, Madingley Road, Cambridge, CB3 0HA, UK}

\begin{document}

\pagerange{\pageref{firstpage}--\pageref{lastpage}} \pubyear{2015}

\maketitle

\label{firstpage}

\begin{abstract}
The distribution of Milky Way halo blue horizontal-branch (BHB) stars is examined using action-based extended distribution functions (EDFs) that describe the locations of stars in phase space, metallicity, and age. 

The parameters of the EDFs are fitted using stars observed in the Sloan Extension for Galactic Understanding and Exploration-II (SEGUE-II) survey that trace the phase-space kinematics and chemistry out to $\sim$70 kpc. A maximum a posteriori probability (MAP) estimate method and a Markov Chain Monte Carlo method are applied, taking into account the selection function in positions, distance, and metallicity for the survey. The best-fit EDF declines with actions less steeply at actions characteristic of the inner halo than at the larger actions characteristic of the outer halo, and older ages are found at smaller actions than at larger actions. In real space, the radial density profile steepens smoothly from $-2$ at $\sim 2$ kpc to $-4$ in the outer halo, with an axis ratio $\sim0.7$ throughout. There is no indication for rotation in the BHBs, although this is highly uncertain. A moderate level of radial anisotropy is detected, with $\beta_s$ varying from isotropic to between $\sim0.1$ and $\sim0.3$ in the outer halo depending on latitude. The BHB data are consistent with an age gradient of $-0.03\,$Gyr kpc$^{-1}$, with some uncertainty in the distribution of the larger ages. These results are consistent with a scenario in which older, larger systems contribute to the inner halo, whilst the outer halo is primarily comprised of younger, smaller systems. 

\end{abstract}


\section[]{Introduction} 
If we assume that the stellar halo is in an approximately steady state, we can characterise it with distribution functions (DFs) $f(\vJ)$ that depend only on the constants of stellar motion $J_i$ \citep{jeans+16}. Using actions as the constants of motion has several clear advantages. First, action coordinates can be complemented by canonically conjugate variables, the angles, to obtain a complete coordinate system for phase space. Second, it is straightforward to add DFs for the thin and thick discs, the bulge and the dark halo to the DF of the stellar halo to build up a complete Galaxy model \citep{piffl+15}. Third, actions are adiabatic invariants and therefore can be used to examine phenomena such as adiabatic contraction \citep[e.g.][]{piffl+15}. Finally, the actions $J_r$, $J_{\phi}$ and $J_z$ quantify excursions of an orbit in the radial, azimuthal and vertical directions, and thus are a natural set of labels for categorizing orbits.

\cite{bell+08} established that much of the halo is comprised of substructures, thought to be relics of disrupted satellites and globular clusters. These clumps disperse in positions and velocities but their ages and metallicities remain bunched \citep{bell+10}. We expect each clump of halo stars sharing the same metallicity and age to have its own DF $f(\mathbf{J},\theta,[\mathrm{Fe/H}],\tau)$. Over time as the clump phase mixes, its DF becomes a function of actions, metallicity, and age only, i.e. an extended distribution function (EDF) of the type introduced by \cite{sanders+15} for the Milky Way disc(s), and developed by \cite{das+16} for the Milky Way halo K giants. The EDF of the entire halo is simply the sum of these individual EDFs. The ages and metallicities of stars are thus associated with separations in action space that may manifest as gradients in real and velocity space.

Several methods have been explored in the literature for determining the distribution of ages of halo stars. These include estimating the main-sequence turn-off temperature and combining it with metallicities and isochrones, finding $10$--$12\,$Gyr \citep{jofre+11}, $10.5 \pm 1.5\,$Gyr \citep{guo+16}, and a minimum age of $8\,$Gyr \citep{hawkins+14}. \cite{kali+12} find an age of $11.4\pm 0.7\,$Gyr for local field halo white dwarf stars, by finding a simple relation between their current and progenitor masses using stellar evolution models. There is some evidence in the literature for an old inner halo with a small dispersion in ages compared to a younger outer halo with a larger dispersion in ages \citep{marquez+94}. \cite{preston+91} and \cite{santucci+15} analyse blue horizontal branch (BHB) stars and find that the mean unreddened colour $B-V$ increases outwards to 40 kpc. Interpreting this as an age gradient amounts to a spread of roughly $2$--$2.5\,$Gyr in age, with the oldest stars concentrated in the central 15 kpc of the Galaxy.

In this work, we revisit the case for a stellar population gradient in the halo using a spectroscopic sample of BHB stars, within the context of EDFs. Following the work of \cite{das+16}, where only a very weak metallicity gradient was found in the K giants, we attribute any gradient in the stellar population to ages. We introduce the EDFs, the initial mass function, and the gravitational potential in Section ~\ref{sec:halomod}. We consider four cases; the first two fit the position and metallicity observables and the second two fit position, velocity, and metallicity observables. In Section ~\ref{sec:data}, we introduce the spectroscopic sample of BHBs and the methods used to select them. In Section ~\ref{sec:fitdata}, the method used to explore the posterior distribution of the observations given the stellar halo model, is described. The observables are a single realization of the convolution of the EDF with the selection function (SF) imposed in selecting the sample. The method of deriving the SF is discussed here. Section ~\ref{sec:results} presents the fits to the observables and properties of our EDFs. Section ~\ref{sec:discuss} compares this work to the literature and assembles an interpretation of the results. We conclude in Section ~\ref{sec:conc}.
\section[]{Stellar halo models}\label{sec:halomod}
An extended distribution function (EDF) gives the probability density of
stars in the space specified by the phase-space coordinates
$(\mathbf{x},\mathbf{v})$ and the variables that characterise stellar properties, such as
mass $m$, age $\tau$, and chemistry ([Fe/H], $\aFe,\ldots$). Below the mass at which the stellar lifetime becomes equal to $\tau$, we assume that the halo's EDF is proportional to the Kroupa IMF \citep{kroupa+93},   
\begin{equation}
	\epsilon(m) = 
	\begin{cases}
		0.035m^{-1.5} &\mathrm{if} \, 0.08 \leq m < 0.5\\
		0.019m^{-2.2} &\mathrm{if} \, 0.5 \leq m < 1.0\\
		0.019m^{-2.7} &\mathrm{if} \, m \geq 1.0\,.
	\end{cases}
\end{equation}
where $m$ is in solar masses. The EDF vanishes at higher masses. Thus the only explicit dependencies of the EDF are on phase-space coordinates, [Fe/H], and age, and the EDF can be considered either a function $f(\mathbf{x},\mathbf{v},\FeH, \tau)$ or a function $f(\mathbf{J},\FeH, \tau)$. For actions we use $J_r, J_\phi\equiv L_z$ and $J_z$.  Where required, we use the St\"{a}ckel Fudge \citep{binney+12} to convert $(\vx,\vv)$ to $\vJ$. We take the gravitational potential to be axisymmetric, and thus cylindrical polar coordinates $(R,\phi,z,v_R,v_{\phi},v_z)$ are a natural choice. 

Below we introduce the EDFs and the gravitational potential. The former includes two `pseudo' EDFs that depend only on positions and metallicities. We include these both as an intermediate step to constructing the full phase-space EDFs and as a means to comparing with past works that only fit density profiles. They are equivalent to full phase-space EDFs integrated over velocities.
\subsection[]{Separable EDF of density and metallicity (constant axis ratio)}\label{ssec:mod1}
The EDF is specified as
\begin{equation}\label{eq:mod1}
{\mathrm{d}N\over\mathrm{d}^3\mathbf{x} \,\mathrm{d}\FeH}
= f(\mathbf{x},G) = Af_{\mathrm{bpl}}(R,z)f_{\mathrm{m}}(G), 
\end{equation}
where $A$ is a normalization constant enforcing a total probability of one and 
\begin{equation}\label{eq:G}
G = -\ln ([\mathrm{Fe/H}]_{\mathrm{max}} - \FeH).
\end{equation}

\noindent The function $f_\mathrm{bpl}$ is the density profile and is specified by a broken power law and a constant axis ratio \citep{deason+12a}
\begin{equation}
f_{\mathrm{bpl}}(R,z)= 
	\begin{cases}
    	\left(\frac{R^2+z^2/q^2}{r_{\mathrm{b}}^2}\right)^{-\alpha_{\mathrm{in}}}/2,& \text{if } R^2+z^2/q^2 \leq r_{\mathrm{b}}^2\\
    	\left(\frac{R^2+z^2/q^2}{r_{\mathrm{b}}^2}\right)^{-\alpha_{\mathrm{out}}/2}, & \text{otherwise}
	\end{cases}
\end{equation}
We consider the metallicity to be a lognormal in $\FeH$ \citep{das+16}
\begin{equation}\label{eq:fm}
	f_{\mathrm{m}}(G) = \e^{G}\frac{\e^{-\frac{G^2}{2\sigma^2}}}{\sigma\sqrt{2\pi}},
\end{equation}
where 
\begin{equation}
	\sigma^2 = -\ln(\FeH_{\mathrm{max}} - \FeH_{\mathrm{peak}}).
\end{equation}
Thus, the distribution is specified by a maximum metallicity, and a metallicity at which the distribution peaks. The difference between these metallicities cannot be greater than 1. A glance at the distribution of metallicities in the observations (Fig.~\ref{fig:obshist}) suggests this is suitable.

Model 1 is thus specified by the parameter set
\begin{equation}
	M_1(q,\alpha_{\mathrm{in}},\alpha_{\mathrm{out}},r_{\mathrm{b}},\FeH_{\mathrm{max}},\FeH_{\mathrm{peak}})
\end{equation}
\subsection[]{Separable EDF of density and metallicity (variable axis ratio)}\label{ssec:mod2}
The EDF is specified as
\begin{equation}\label{eq:mod2}
{\mathrm{d}N\over\mathrm{d}^3\mathbf{x} \,\mathrm{d}\FeH}
= f(\mathbf{x},G) = Af_{\mathrm{spl}}(R,z)f_{\mathrm{m}}(G), 
\end{equation}
where $A$ is a normalization constant enforcing a total probability of one. $G$ and $f_m$ are defined by Equations ~\eqref{eq:G} and ~\eqref{eq:fm}, respectively. $f_\mathrm{spl}$ is the density profile and is specified by a single power law with a variable axis ratio \citep{xue+15}
\begin{equation}
f_{\mathrm{spl}}(R,z) = (R^2+z^2/q(r)^2)^{-\alpha/2},
\end{equation}
where
\begin{equation}
\begin{split}
r &= (R^2+z^2)^{-1/2}\\
q(r) &= q_{\infty} - (q_{\infty} - q_0)\exp\left(1 - \frac{\sqrt{r^2+r_0^2}}{r_0}\right).
\end{split}
\end{equation}
Model 2 is thus specified by the parameter set
\begin{equation}
	\
    M_2(q_{\infty},q_0,\alpha,r_0,\FeH_{\mathrm{max}},\FeH_{\mathrm{peak}})
\end{equation}
\subsection[]{Separable EDF of phase space and metallicity}\label{ssec:mod3}

The EDF is specified as
\begin{equation}\label{eq:mod3}
	{\mathrm{d}N\over\mathrm{d}^3\mathbf{x}\,\mathrm{d}^3\mathbf{v}\,\mathrm{d}\FeH} \ = f(\mathbf{J},G) = Af_{\mathrm{ps}}(\mathbf{J})f_{\mathrm{m}}(G),
\end{equation}
where $A$ is a normalization constant enforcing a total probability of one. $f_{\mathrm{m}}$ is given by Equation \eqref{eq:fm}, and $f_{\mathrm{ps}}$ is from \cite{posti+15}
\begin{equation}\label{eq:postiDF}
	\begin{split}
		f_{\mathrm{ps}}(\mathbf{J})  &=  \frac{[1+J_0/h(\mathbf{J})]^{\beta_{\mathrm{in}}}}{[1+g(\mathbf{J})/J_0]^{\beta_{\mathrm{out}}}}\\
		h(\mathbf{J}) 				&= a_rJ_r + a_{\phi}|J_{\phi}| + a_zJ_z\\
		g(\mathbf{J})  				&= b_rJ_r + b_{\phi}|J_{\phi}| + b_zJ_z.
	\end{split}
\end{equation}

\noindent For $|\vJ|\gg J_0$, $f_{\rm ps}$ is dominated by $g(\vJ)$, and for $|\vJ|\ll J_0$, $f_{\rm ps}$ is
dominated by $h(\mathbf{J})$. Both $g(\vJ)$ and $h(\vJ)$
are homogeneous functions of the actions of degree one.

The parameters $(a_r,a_{\phi},a_z,b_r,b_{\phi},b_z)$ control the shape of the density and velocity ellipsoids. Rescaling the $a_i$ and $b_i$ by the same factor has no effect on the model if accompanied by a rescaling of $J_0$. This degeneracy is eliminated by imposing the conditions $\sum_i a_i=\sum_i b_i=3$. 

Model 3 is thus specified by 
\begin{equation}
M_3(\beta_{\mathrm{in}},\beta_{\mathrm{out}},J_0,a_r,a_{\phi},b_r,b_{\phi},\FeH_{\mathrm{max}},\FeH_{\mathrm{peak}}).
\end{equation}
\subsection[]{Correlated EDF of phase space, metallicity, and age, with rotation}\label{ssec:mod4}
The EDF is specified as
\begin{equation}\label{eq:mod4}
	\begin{split}
	{\mathrm{d}N\over\mathrm{d}^3\mathbf{x}\,\mathrm{d}^3\mathbf{v}\,\mathrm{d}\FeH\,\mathrm{d}\tau} &= f(\mathbf{J},G,\tau) \\
    							  &= Af_{\mathrm{psr}}(\mathbf{J})f_\mathrm{m}(G)f_{\mathrm{psa}}(\mathbf{J},\tau),
	\end{split}
\end{equation}
where $A$ and $f_{\mathrm{m}}$ are as above. $f_{\mathrm{psr}}$ is given by
\begin{equation}\label{eq:psr}
	\begin{split}
		f_{\mathrm{psr}}(\mathbf{J}) &= R(J_{\phi})f_{\mathrm{ps}}(\mathbf{J})\\
	    R(J_{\phi}) & = 1 + x\tanh\left(\frac{J_{\phi}}{J_0}\right),	
    \end{split}
\end{equation}
where $f_{\mathrm{ps}}$ is given by Equation \eqref{eq:postiDF}. The prefactor $R(J_{\phi})$ splits the phase-space DF into even and odd components, introducing the possibility for rotation. $x$ governs the strength of the rotation. $f_{\mathrm{psa}}$ is given by
\begin{equation}\label{eq:psa}
	f_{\mathrm{psa}} = \delta\left(\tau - \left[a_{\tau} + b_{\tau}\ln{\frac{J_t}{J_0}}\right]\right), 
\end{equation}
where $J_{\mathrm{t}}$ is the total action
\begin{equation}
J_{\mathrm{t}} = \sqrt{J_r^2+J_{\phi}^2 + J_z^2}.
\end{equation}
\noindent This implies a single age at each total action, which is a guide to apocentric radius. $b_{\tau}$ encodes the dependence on actions, and $a_{\tau}$ is the age for which the total action is equal to the transition action $J_0$. Increasing $|b_{\tau}|$ increases the age gradient within the halo, with $b_{\tau} < 0$ implying that mean age decreases with radius. Increasing $a_{\tau}$ makes the halo older at every radius.

Model 4 is specified by
\begin{equation}
M_4(\beta_{\mathrm{in}},\beta_{\mathrm{out}},J_0,a_r,a_{\phi},b_r,b_{\phi},
x,\FeH_{\textrm{max}},\FeH_{\textrm{peak}},a_{\tau},b_{\tau}).
\end{equation}
\begin{table}
 \centering
  \caption{Parameters of the Galactic potential.\label{tab:potpars}}
  \begin{tabular}{lll}
  	\hline
  	Component     		&Parameter     					 &Value\\
  	\hline
  	Thin          		&$R_\mathrm{d}$ (kpc)     				 &2.682\\
  					    &$z_\mathrm{d}$ (kpc)     				 &0.196\\
  					    &$\Sigma_\mathrm{d} (M_{\odot}$kpc$^{-2}$)  &5.707$\times10^8$\\
  	\hline
  	Thick       			&$R_\mathrm{d}$ (kpc)     				 &2.682\\
  					    &$z_\mathrm{d}$ (kpc)     				 &0.701\\
  					    &$\Sigma_\mathrm{d} (M_{\odot}$kpc$^{-2}$)  &2.510$\times10^8$\\
  	\hline
  	Gas         			&$R_\mathrm{d}$ (kpc)     				 &5.365\\
  					    &$z_\mathrm{d}$ (kpc)     				 &0.040\\
  					    &$\Sigma_\mathrm{d} (M_{\odot}$kpc$^{-2}$)  &9.451$\times10^7$\\
  					    &$R_{hole}$ (kpc)				 &4.000\\
  	\hline
  	Bulge	  			&$\rho_0 (M_{\odot}$kpc$^{-3}$)    &9.490$\times10^{10}$\\
  						&$q$								 &0.500\\
  						&$\gamma$						 &0.000\\
  						&$\delta$						 &1.800\\
  						&$r_0$ (kpc)						 &0.075\\
  						&$r_\mathrm{t}$ (kpc)					     &2.100\\
  	\hline
  	Dark halo    &$\rho_0 (M_{\odot}$kpc$^{-3}$)    &1.815$\times10^7$\\
  						&$q$								 &1.000\\
  						&$\gamma$						 &1.000\\
  						&$\delta$						 &3.000\\
  						&$r_0$ (kpc)						 &14.434\\
  						&$r_\mathrm{t}$ (kpc)
						&$\infty$\\
  	\hline
  \end{tabular}
\end{table}
\subsection[]{The gravitational potential}
As in previous works, we use the composite potential proposed by \cite{dehnen+98}, generated by thin and thick stellar discs, a gas disc, and two spheroids representing the bulge and the dark halo. The densities of the discs are
given by
\begin{equation}
	\rho_\mathrm{d}(R,z) = \frac{\Sigma_0}{2z_\mathrm{d}}\exp\left[-\left(\frac{R}{R_\mathrm{d}} + \frac{|z|}{z_\mathrm{d}} + \frac{R_{\mathrm{hole}}}{R}\right) \right],
\end{equation}
where $R_\mathrm{d}$ is the scale length, $z_\mathrm{d}$, is the scale height, and
$R_{\mathrm{hole}}$ controls the size of the hole at the centre of the disc,
which is only non-zero for the gas disc. The densities of the bulge and dark
halo are given by
\begin{equation}
	\rho(R,z) = \rho_0\frac{(1+m)^{(\gamma-\delta)}}{m^{\gamma}}\,\exp\left[-(mr_0/r_{\mathrm{t}})^2\right],
\end{equation}
where 
\begin{equation}
	m(R,z) = \sqrt{(R/r_0)^2 + (z/qr_0)^2}. 
\end{equation}
$\rho_0$ sets the density scale, $r_0$ is a scale radius, and the parameter $q$ is the axis ratio of the isodensity surfaces. The exponents $\gamma$ and $\delta$ control the inner and outer slopes of the radial density profile, and $r_\mathrm{t}$ is a truncation radius. 

The adopted parameter values are taken from \cite{piffl+14} and given in Table~\ref{tab:potpars}. They specify a spherical NFW halo that is not truncated
($r_\mathrm{t}=\infty$). The stellar halo contributes only
negligible mass, and thus can be considered included in the contributions of the bulge and dark halo. 

\section[]{Observational constraints}\label{sec:data}
\begin{figure*}
	\centering
	\includegraphics[scale=0.46]{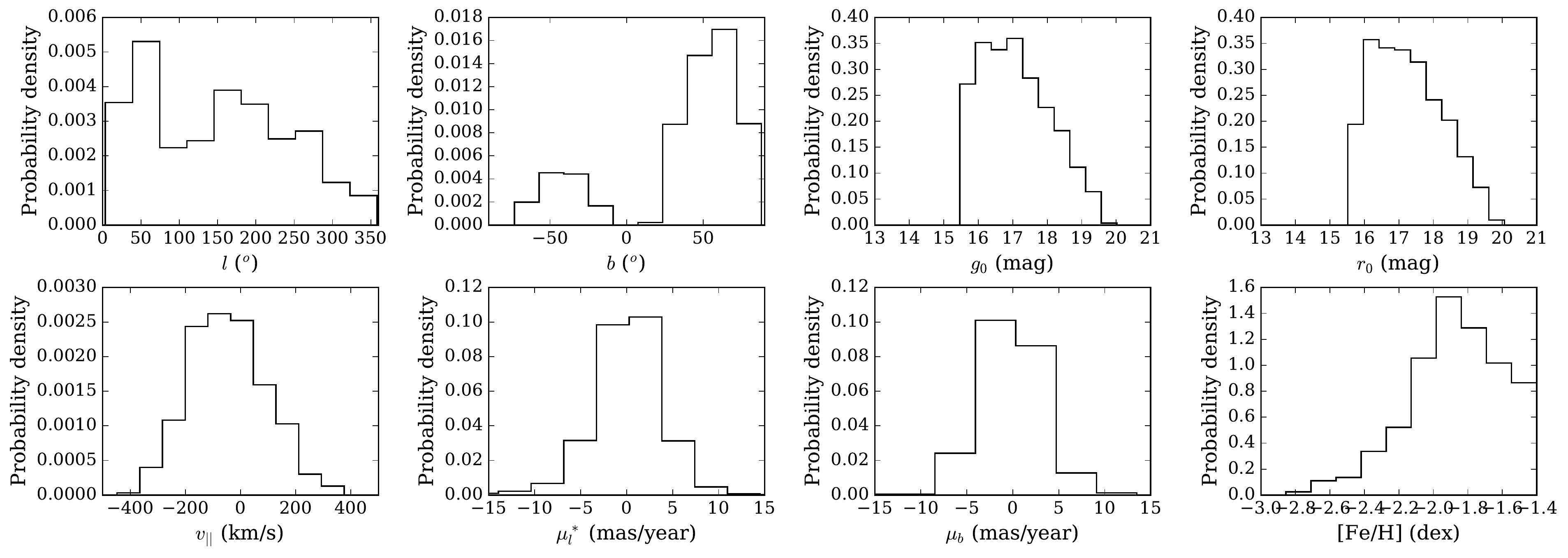}
	\caption{One-dimensional distributions for the SEGUE-II BHBs in Galactic coordinates, for sky positions, apparent magnitudes, line-of-sight velocities, proper motions, and metallicities.\label{fig:obshist}}
\end{figure*}
Here, we introduce the observations that we will use to constrain parameters
of the EDFs. We adopt a left-handed coordinate system in which positive $v_R$ is away from the Galactic centre and positive $v_{\phi}$ is in the direction of Galactic rotation. To convert from Galactocentric coordinates to heliocentric coordinates we assume that the Sun is located at $(R_0,z_0) =
(8.3,0.014)\kpc$ \citep{schonrich+12,binney+97}, that the local standard of
rest (LSR) has an azimuthal velocity of $238\kms$, and that the velocity of
the Sun relative to the LSR is $(v_R,v_{\phi},v_z) = (-14.0, 12.24,
7.25)\kms$ \citep{schonrich+12}.

\subsection[]{BHB sample}
We use the BHB sample of \cite
{xue+11}, which is based on Data Release 10 of the Sloan Digital Sky Survey (SDSS) and in particular the Sloan Extension for Galactic Understanding and Exploration surveys (SEGUE) within it. The sample consists of equatorial coordinates $(\alpha,\delta)$, apparent magnitudes, colours, line-of-sight velocities $v_{||}$, and spectroscopic metallicities $\FeH$. We supplement the catalogue values with proper motions ($\mu_l^* = \dot{l}\cos b$ and $\mu_b = \dot{b}$)  downloaded from SkyServer's CasJobs\footnote{\url{http://skyserver.sdss.org/CasJobs/}}, by cross matching within 15 arcsec. We include proper motion measurements to use all available data, but note that the uncertainties are often $> 100\%$ and so we do not expect much, if any, extra constraining power to come from their inclusion. We constrain the sample to SEGUE-II stars only, as the selection criteria are not fully known for SDSS Legacy and SEGUE-I stars. SEGUE II focuses on distant stars and therefore only uses `faint' plates. SEGUE I used both `bright' and `faint' plates to cover a larger range in apparent magnitudes. We remove stars on cluster and test
plates and apply the cut $\FeH\le-1.4$ to reduce the contamination from
disc stars \citep{schonrich+12}. We exclude stars on plates that intersect the two polygons given by \cite{fermani+13a} as containing the Sagittarius stream. We use the relation of \cite{fermani+13a} to relate apparent magnitudes, colours, and metallicity to a
heliocentric distance\footnote{We shorten `heliocentric distance' to
distance for the remainder of the paper.} $s$ for the BHBs. Thus our observables are defined by the vector
\begin{equation}\label{eqn:obs}
	\mathbf{u} = (l,b,s,v_{||},\mu_l^*,\mu_b,\FeH).
\end{equation}
\begin{table*}
 \centering
  \caption{Combined selection criteria in terms of apparent magnitude, colour indices, and metallicity. $n^*$ is the number of stars after applying the combined selection criteria, and $u$, $g$, $r$, and $i$ refer to apparent magnitudes in Sloan's $ugriz$ colour-magnitude system. \label{tab:selfunc}}
  \begin{tabular}{lllll}
  	\hline
	Programme 	&$n^*$    &Apparent  	&Colour				     &Metallicity\\	
		 		&	 	&magnitude 		&					     &\\	
	\hline		
	SEGUE II 	&701	 &$15.5<g<19.0$	&$0.8<(u-g)<1.5$		 &$\text{[Fe/H]}<-1.4$\\
	 			& 		 &$r>12.5$		&$-0.4<(g-r)<0.0$		 &\\
	\hline	
	\end{tabular}
\end{table*}
\subsection[]{Selection criteria}
Our selection function is the overlap between the original spectroscopic
targeting criteria, criteria imposed by \cite{xue+11}, and further criteria applied by us. The selection on sky positions is given by the coverage of the spectroscopic plates on the sky, and on each plate by the completeness of the spectroscopic sample, i.e. the number of stars observed compared to the potential number of targets. The remaining combined selection criteria relate to apparent magnitudes and various colour indices. These are summarised in Table~\ref{tab:selfunc}. The phase-space, colour, and metallicity distributions for the sample are shown in Fig.~\ref{fig:obshist}. 
\section[]{Fitting the data}\label{sec:fitdata}
We construct the likelihood of models by the method of \cite{mcmillan+13} that  \cite{das+16} used to fit an EDF to halo K giants. The form and contributing terms of the likelihood are described in detail here. 
\subsection[]{The likelihood from Bayes' law}
The total likelihood $\mathcal{L}$ of a model $M$ is given by the product over all stars $i$ of the individual likelihoods $\mathcal{L}^i=P(\mathbf{u}^i|SM)$ of
measuring the star's catalogued coordinates $\mathbf{u}^i$ given the model $M$ and that it is in the survey $S$. By Bayes' law this is
\begin{equation}	
		\mathcal{L}^i = \frac{P(S|\mathbf{u}^i)P(\mathbf{u}^i|M)}{P(S|M)}.
\end{equation}
\noindent $P(S|\mathbf{u}^i)$ is the probability that the star is in the survey given the observables i.e. the `selection
function' (Section ~\ref{ssec:selfunc}). $P(\mathbf{u}^i|M)$ is the EDF convolved with the error distribution of the observables (Section ~\ref{ssec:errconv}). $P(S|M)$ is 
the probability that a randomly chosen star in the model enters the catalogue (Section ~\ref{ssec:normfactor}). The total log-likelihood is
\begin{multline}\label{eqn:logL}
	\log \mathcal{L} = \sum_{i=k}^{n_*} \log \mathcal{L}^i = 
	 \sum_{i=k}^{n_*}\log\left(P(S|\mathbf{u}^i)\right) 
+\\
\hskip2cm \sum_{i=k}^{n_*}\log\left(P(\mathbf{u}^i|M)\right) - n_*\log\left(P(S|M)\right),
\end{multline}
where $n_*$ is the number of stars. 
\subsection[]{Selection function}\label{ssec:selfunc}
There is no selection on line-of-sight velocities or proper motions. Moreover, we assume that the selection function is separable as
\begin{equation}
	\begin{split}
		P(S|\mathbf{u}^i) =& \,p(S|l,b,s,\mathrm{[Fe/H]})\\
	   			      	  =& \,p(S|l,b)\,p(S|s,\mathrm{[Fe/H]}).
	\end{split}
\end{equation}
The selections on sky positions $p(S|l,b)$ and distance/metallicity $p(S|s,\mathrm{[Fe/H]})$ are described below.
\subsubsection[]{Selection on sky positions}\label{sssec:skypossf}
\begin{figure}
\centering
\includegraphics[scale=0.6]{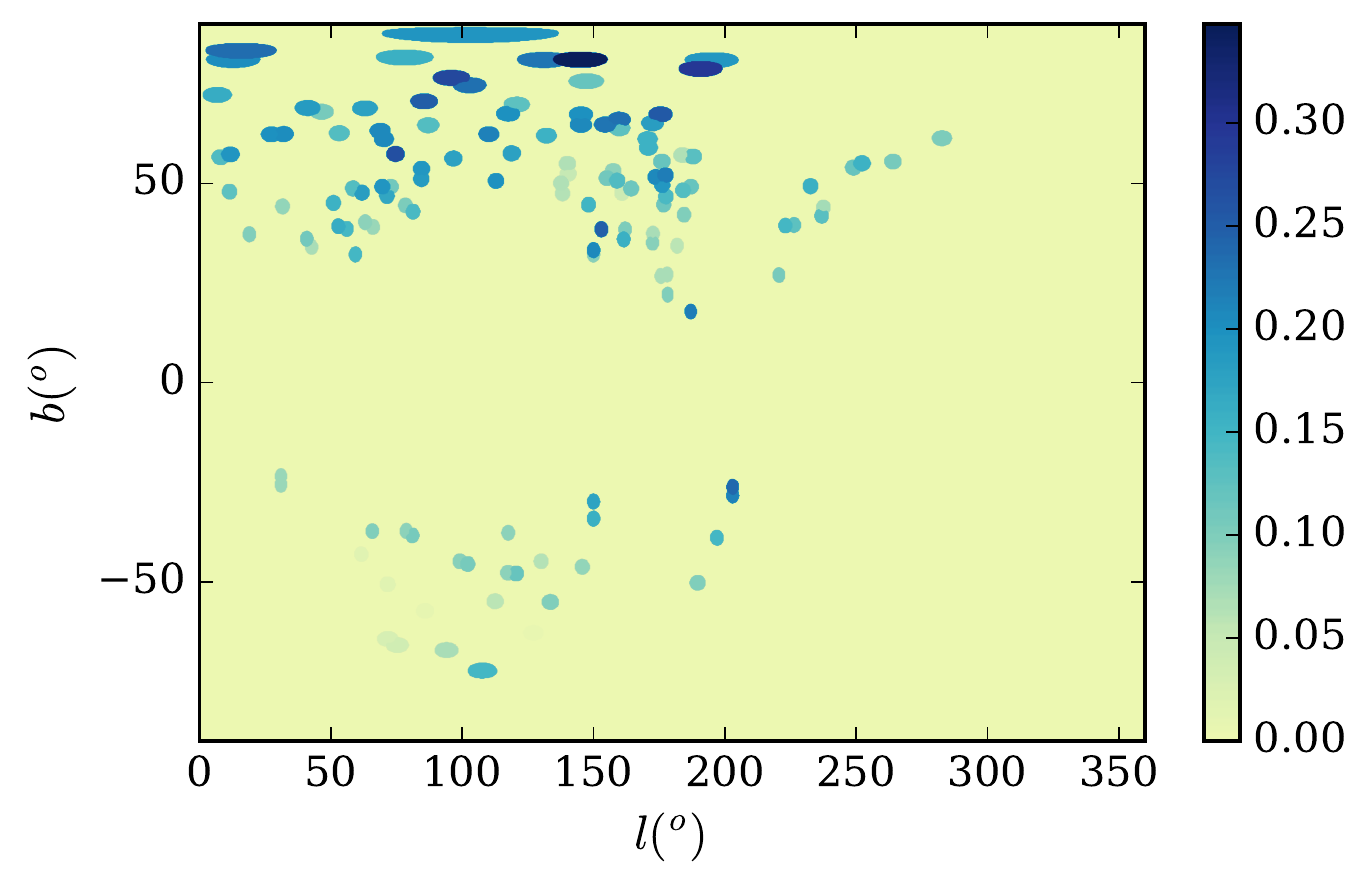}
 \caption{Selection function as a function of sky positions,  $p(S|l,b)$.\label{fig:lbsfmaps}}
\end{figure}
The selection on $l$ and $b$, $p(S|l,b)$, depends on the coordinates of the SEGUE-II plates and the completeness fraction, i.e. the fraction of photometrically identified targets for which spectra were obtained. The completeness fraction depends strongly on $|b|$ because close to the plane available targets are numerous so an individual star has a low probability of being allocated a fibre. For coordinates within $1.49\degree$ of the centre of a plate, the selection function equals the completeness fraction for that plate. Thus
\begin{equation}
	P(S|l,b) = \frac{N_{\mathrm{spec,plate}}}{N_{\mathrm{phot,plate}}},
\end{equation}
where $N_{\mathrm{spec,plate}}$ is the number of BHBs on the plate that make the spectroscopic sample and $N_{\mathrm{phot,plate}}$ is the number of BHBs in the photometric sample within the same patch in the sky. We evaluate these fractions by searching for BHBs in the spectroscopic and photometric samples in the regions covered by each of the plates, using SkyServer's CasJobs. In Fig.~\ref{fig:lbsfmaps} the colour scale shows $p(S|l,b)$. The dependence of $p(S|l,b)$ on $b$ is evident.
\begin{figure*}
\centering
	\subfloat[]{
		\includegraphics[scale=0.4]{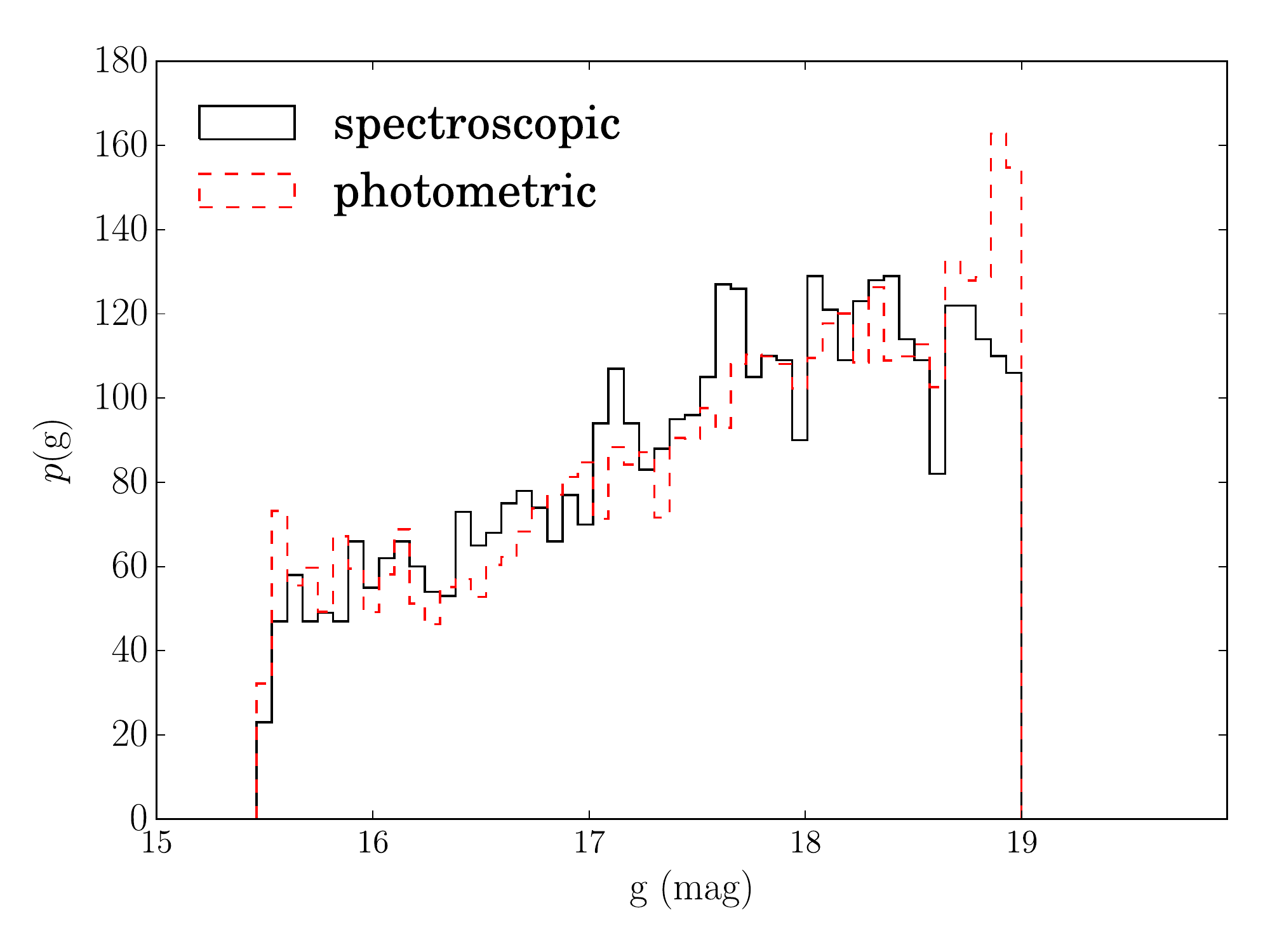}
	}
	\subfloat[]{
		\includegraphics[scale=0.4]{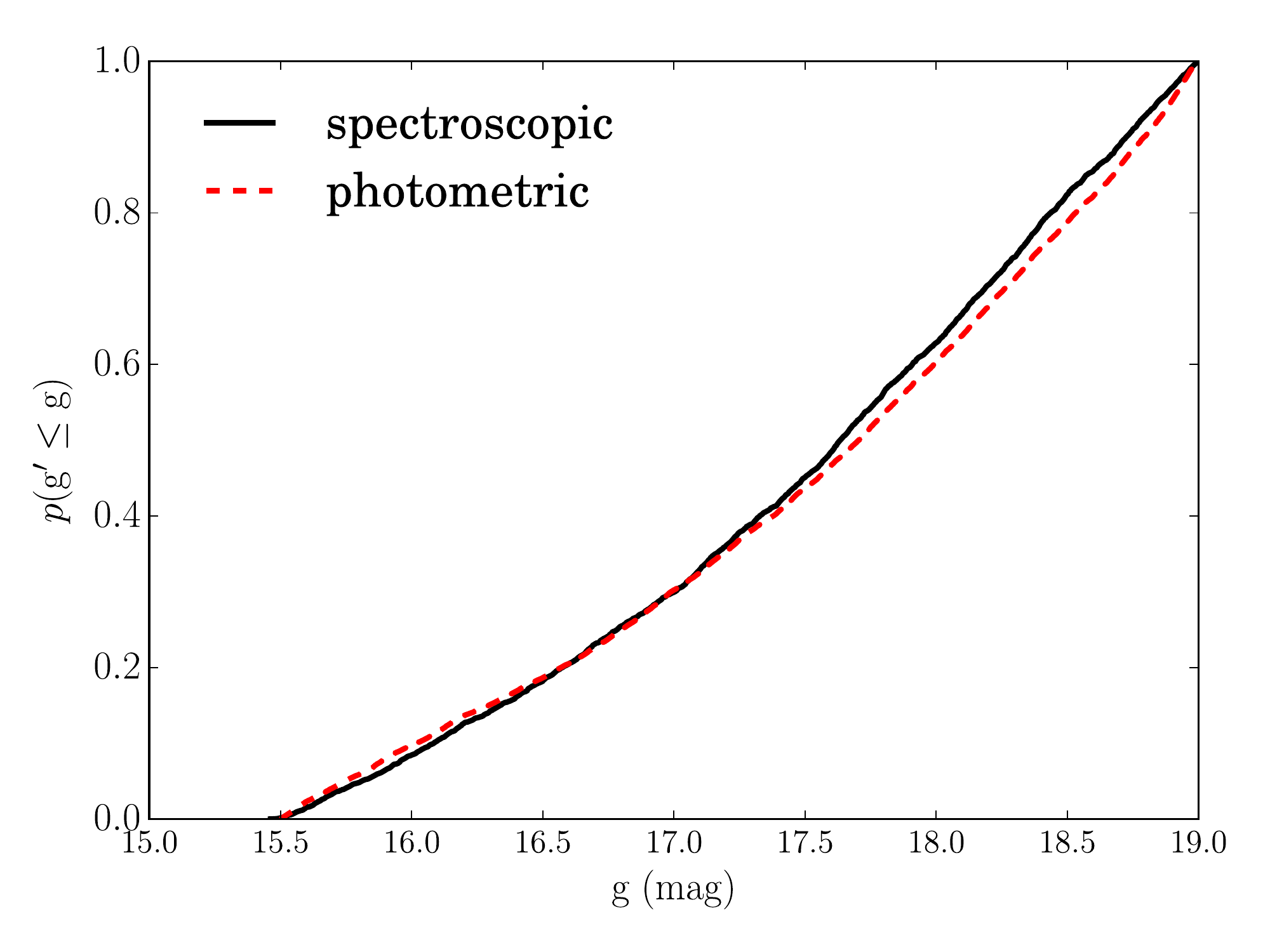}
}
\caption{Probability (a) and cumulative (b) $g$-band apparent magnitude distributions for the photometric and spectroscopic samples.\label{fig:magdist}}
\end{figure*}
\begin{figure}
\centering
q\includegraphics[scale=0.5]{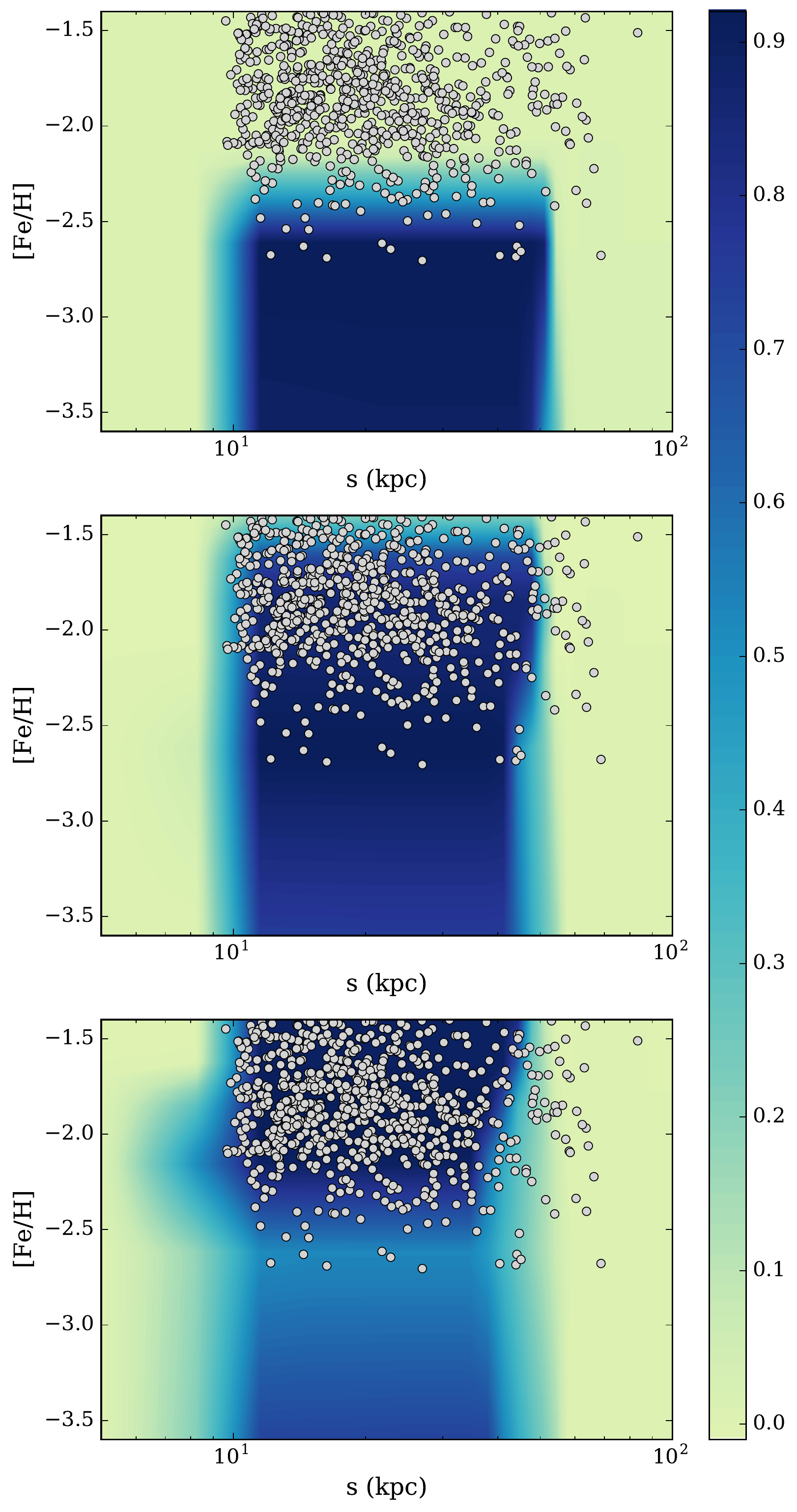}
 \caption{Selection function as a function of distance and metallicity, assuming a single age of $9$ (top), $11$ (middle), and $13$ (bottom) Gyr, with locations of observed stars superimposed.\label{fig:chemsfmaps_singleage}}
\end{figure}
\subsubsection[]{Selection on distance and metallicity}
The selection probability in terms of distance and metallicity depends on the assumed IMF, the survey selection with respect to apparent magnitudes and colours, and the isochrones used to relate apparent magnitudes and colours to intrinsic properties. We assume the survey selection on colour to be uniform over the ranges in Table~\ref{tab:selfunc}. The signal-to-noise ratio however decays with apparent magnitude, and therefore the survey selection is not uniform within the imposed apparent magnitude range. Fig.~\ref{fig:magdist} shows the $g$-band apparent magnitude distribution for our sample of SEGUE-II stars and for stars in the SDSS photometric sample found in Section ~\ref{sssec:skypossf} to lie within the plate dimensions and the selection criteria of Table~\ref{tab:selfunc}. The left panel shows the distributions normalized to unity for $g$-band apparent magnitudes between $15.5$ and $19.0$, just within the range imposed by the SEGUE-II targetting criteria for BHBs ($15.5<g<20.3$). The right panel shows the associated cumulative distribution function. The plots show that the single faint SEGUE-II plates do an excellent job of capturing the apparent magnitude distribution out to a $g$-band apparent magnitude of $19.0$. Therefore we take the selection on apparent magnitude to be uniform within the ranges specified in Table~\ref{tab:selfunc} and zero otherwise.

We use the $\alpha$-enhanced B.A.S.T.I. isochrones for a mass-loss parameter $\eta=0.4$ \citep{pietrin+06} to relate apparent magnitudes to intrinsic stellar properties in the sample. We tabulate on a grid in $s$ and [Fe/H] the probability that a star randomly chosen at birth passes the magnitude and colour cuts
\begin{equation}
p(S|\tau,s,\FeH) = \int \diff m \, \epsilon(m) p(S|m,\tau,s,\FeH).
\end{equation}
For a given value of $s$, we integrate each isochrone (specified by [Fe/H] and $\tau$) over the mass distribution, by adding a star's IMF value to $p(S|s,\FeH,\tau)$ only if it passes the magnitude and colour cuts. We repeat this for a grid of distances, and create an interpolant of $p(S|s,\FeH,\tau)$ that gives the selection function probability for any distance, metallicity, and age within the specified domain. 

The colour scale in Fig.~\ref{fig:chemsfmaps_singleage} shows $p(S|s,\FeH,\tau)$ for $\tau = 9, \, 11$, and $13$ Gyr. The top plot shows that the BHBs cannot be described by a single age of $9\,$Gyr, as stars of metallicities higher than $-2.3$ would then not be seen. However the stars could be described by a single age of $11$ or $12\,$Gyr, i.e. all observed stars lie within the boundaries of non-negligible probabilities predicted for these single ages. We consider two possible age distributions for the BHBs. The first is a delta function, centred on $11\,$Gyr \citep{jofre+11,das+16}, and the second is a function of the actions, as prescribed by Equation \eqref{eq:psa}. 

For the first case, a glance at Fig.~\ref{fig:chemsfmaps_singleage} shows that $p(S|s,\FeH,\tau=11)$ is approximately constant within well-defined boundaries. Therefore to avoid the time-consuming engagement with the isochrones, we assume a constant probability within a box, so 
\begin{equation}\label{eqn:chemsfp_const}
p(S|s,\FeH) =
	\begin{cases}
    	\multirow{2}{*}{1}   & \mathrm{if}\,10\, \mathrm{kpc}\leq s\leq 50\, \mathrm{kpc}\\
                              & \mathrm{and}\,-3.6\leq\FeH\leq -1.4\\
		0 &\, \mathrm{otherwise.}\\
	\end{cases}
\end{equation}
We adopt this age distribution for the models specified in Sections \ref{ssec:mod1}, \ref{ssec:mod2}, and \ref{ssec:mod3} ($M_1$, $M_2$, and $M_3$). For the model described in Section \ref{ssec:mod4} ($M_4$), we assume the age to depend on actions within the same distance and metallicity limits as specified in Equation ~\eqref{eqn:chemsfp_const}.

\subsection[]{Convolution of the EDF with the error distribution}\label{ssec:errconv}
We neglect errors in sky coordinates and adopt independent Gaussian error
distributions for $v_\parallel$, $\mu_l^*$, $\mu_b$, [Fe/H], and $\log s$.
Thus the multi-variate error distribution is
\begin{equation}\label{eqn:convint}
	C^{(7)}(\mathbf{u}^i,\mathbf{u}^{\prime},\mathbf{p}^i) \equiv
	\prod_{l=1}^2\delta(u_l^i-u_l')\prod_{l=3}^{7}C(u_l^i,u_l^{\prime},p_l^i),
\end{equation}
where true values of observables are indicated by primes and
\begin{equation}
C(u^i,u',p^i)\equiv{\exp\left[{-(u^i-u')^2/(2p^{i2})}\right]\over\sqrt{2\pi} p^i}.
\end{equation}
The
convolution of the EDF with the error distribution of a given star is
\begin{multline}
	P(\mathbf{u}^i|M) = \\
	\int \diff^7\mathbf{u}^{\prime} \, C^7(\mathbf{u}^i,\mathbf{u}^{\prime},\mathbf{p}^i) \, f(\mathbf{\mathbf{x},\mathbf{v},\FeH})) \left|\frac{\partial(\mathbf{x},\mathbf{v},\FeH)}{\partial(\mathbf{u}^{\prime})}\right| .
\end{multline}
The Jacobian determinant here is proportional to $s^4\cos b$
\citep{mcmillan+12}. The
integral is calculated using a fixed Monte Carlo sample of 2000 points per star to eliminate Poisson noise in likelihood evaluations \citep{mcmillan+13}.
\begin{table}
 \centering
  \caption{Median and 68$\%$ confidence intervals for $M_1$, $M_2$, and $M_4$, and MAP estimates for $M_3$. A `*' indicates that the parameter is either set prior to the runs, or fixed by other parameters. \label{tab:bestfitpars}}
\begin{tabular}{lllll}
  	\hline
  	Parameter				&$M_1$	 					&$M_2$						&$M_3$			&$M_4$\\
    \hline
    $q$						&$0.72_{-0.030}^{+0.031}$	&-							&-				&-\\
    $\alpha_{\mathrm{in}}$	&$3.61_{-0.16}^{+0.15}$		&-							&-				&-\\
    $\alpha_{\mathrm{out}}$	&$4.75_{-0.28}^{+0.30}$		&-							&-				&-\\
	$r_{\mathrm{b}}$/kpc	&$29.87_{-3.55}^{2.80}$	    &-							&-				&-\\ 
    \hline
    $q_{0}$					&-							&$0.39_{-0.09}^{+0.08}$		&-				&-\\
    $q_{\infty}$			&-							&$0.81_{-0.050}^{+0.055}$	&-				&-\\
    $r_0$/kpc				&-							&$7.32_{-1.73}^{+1.88}$		&-				&-\\
    $\alpha$				&-							&$4.65_{-0.23}^{+0.25}$		&-				&-\\
    \hline
    $\beta_{\mathrm{in}}$				&-				 			&-							&$2.05$			&$2.17_{-0.52}^{+0.44}$\\
    $\beta_{\mathrm{out}}$				&-				 			&-							&$4.72$			&$4.65_{-0.31}^{+0.33}$\\	    $J_0$     				&-				 			&-							&$1635$			&$1635*$\\
    $a_r$					&-				 			&-							&$1.31$			&$0.70_{-0.24}^{+0.27}$\\
    $a_{\phi}$				&-				 			&-							&$0.93$			&$0.88_{-0.22}^{+0.25}$\\
    $a_z$					&-				 			&-							&$0.78*$		&$1.42*$\\
    $b_r$					&-				 			&-							&$1.11$			&$1.33_{-0.15}^{+0.17}$\\
    $b_{\phi}$				&-				 			&-							&$0.60$			&$0.54_{-0.09}^{+0.10}$\\
    $b_z$					&-				 			&-							&$1.29*$		&$1.13*$\\
    $x$						&-				 			&-							&$0.34$			&$0.07_{-0.16}^{+0.19}$\\
    \hline
    $\FeH_{\mathrm{max}}$   &$-0.83_{-0.017}^{+0.018}$	&$-0.83_{-0.017}^{+0.018}$	&$-0.83_{-0.017}^{+0.018}$		&$-0.80_{-0.030}^{+0.053}$\\
    $\FeH_{\mathrm{peak}}$	&$-1.77_{-0.018}^{+0.020}$	&$-1.77_{-0.018}^{+0.020}$	&$-1.77_{-0.018}^{+0.020}$		&$-1.73_{-0.035}^{+0.059}$\\
    \hline
 	$a_{\tau}$				&-				 			&-							&-				&$12.00_{-0.29}^{+0.37}$\\
    $b_{\tau}$			    &-				 			&-							&-				&$-0.69_{-0.14}^{+0.24}$\\
\hline
  \end{tabular}
\end{table}
\subsection[]{The normalization factor}\label{ssec:normfactor}
The normalization of $\mathcal{L}$ is given by
\begin{multline}
P(S|M) = \\
\int\diff^7\mathbf{u}^{\prime}\,P(S|\mathbf{u}^{\prime})\,f(\mathbf{x},\mathbf{v},\FeH)\left|\frac{\partial(\mathbf{x},\mathbf{v},\FeH}{\partial(\mathbf{u}^{\prime})}\right| .
\end{multline}
We approximate the integrals over sky coordinates by sums of the remaining five-dimensional integrals evaluated at the centre of
each SEGUE-II plate \citep{das+16}. Since $P(S|\mathbf{u}^{\prime})$ is multiplied by  $n_*$ in Equation (\ref{eqn:logL}), the integrals must be calculated to a high degree of accuracy \citep{mcmillan+12}. We calculate them to an accuracy of 0.1\% using a Python wrapper for the cubature method\footnote{This wrapper can be downloaded from \url{https://github.com/saullocastro/cubature}.}. The procedure is parallelised over 16 cores using the distributed memory tool in the \texttt{multiprocessing} package of Python.
\begin{figure*}
	\centering
	\includegraphics[scale=0.4]{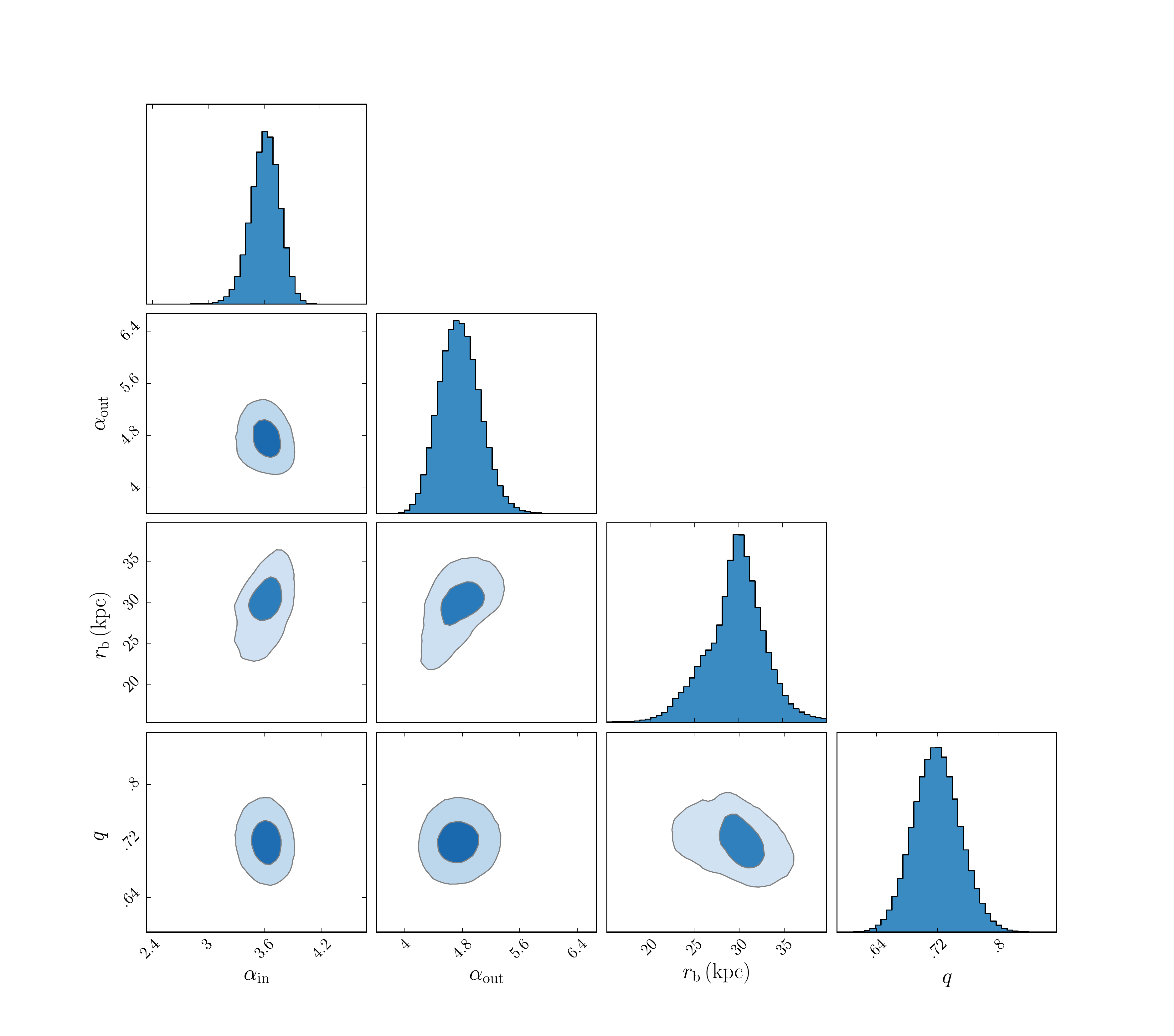}
	\caption{{\tt emcee} results for fitting the DF $M_1$. The plots along the diagonal illustrate 1-D probability distributions for each of the model parameters. The remaining plots show joint probability distributions between all pairs of parameters. The contours show the 1-$\sigma$ and 2-$\sigma$ confidence levels. \label{fig:emcee_m1}}
\end{figure*}
\begin{figure*}
	\centering
	\includegraphics[scale=0.4]{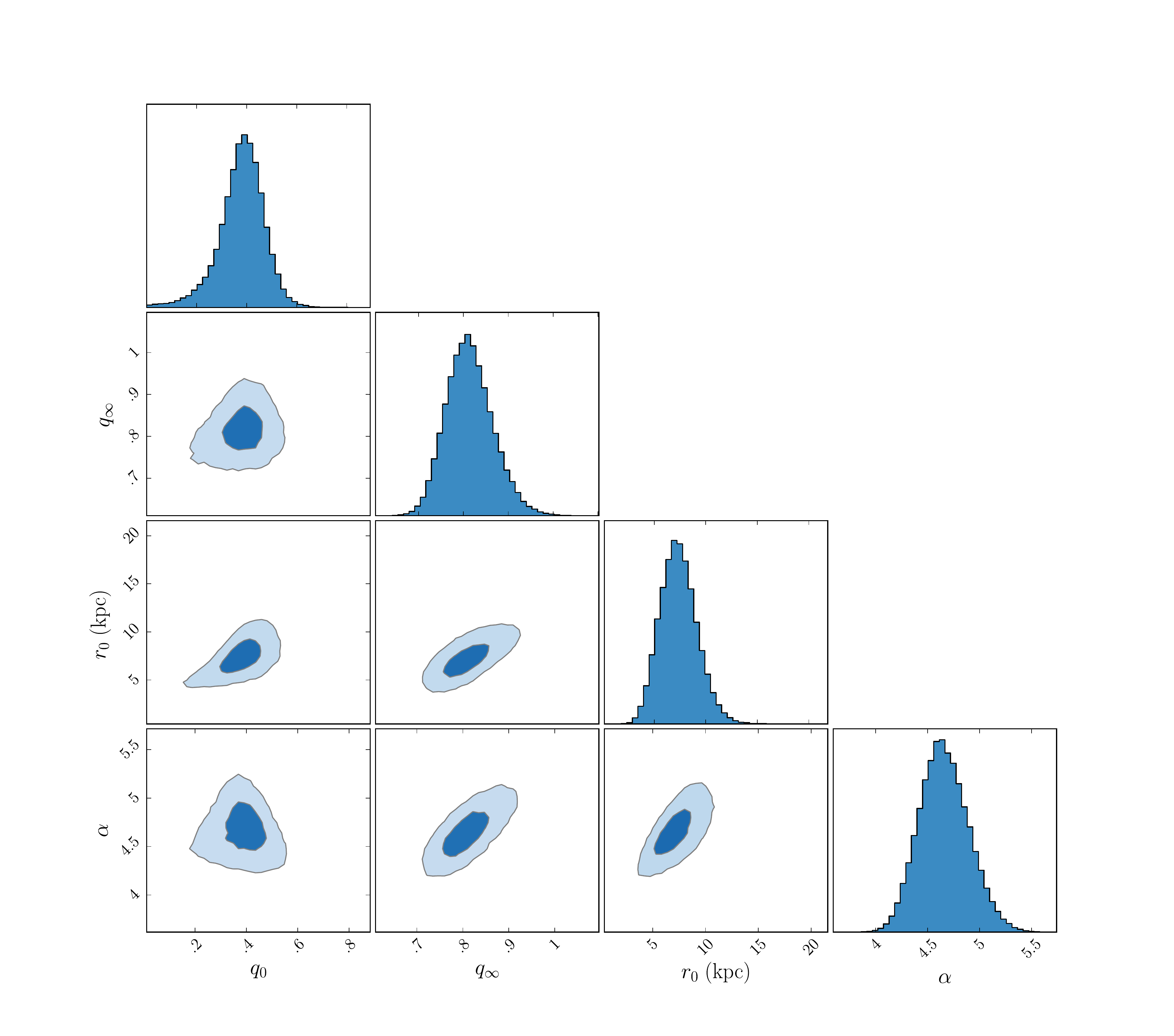}
	\caption{{\tt emcee} results for fitting the DF $M_2$. The plots along the diagonal illustrate 1-D probability distributions for each of the model parameters. The remaining plots show joint probability distributions between all pairs of parameters. The contours show the 1-$\sigma$ and 2-$\sigma$ confidence levels.\label{fig:emcee_m2}}
\end{figure*}
\begin{figure}
	\centering
	\includegraphics[scale=0.25]{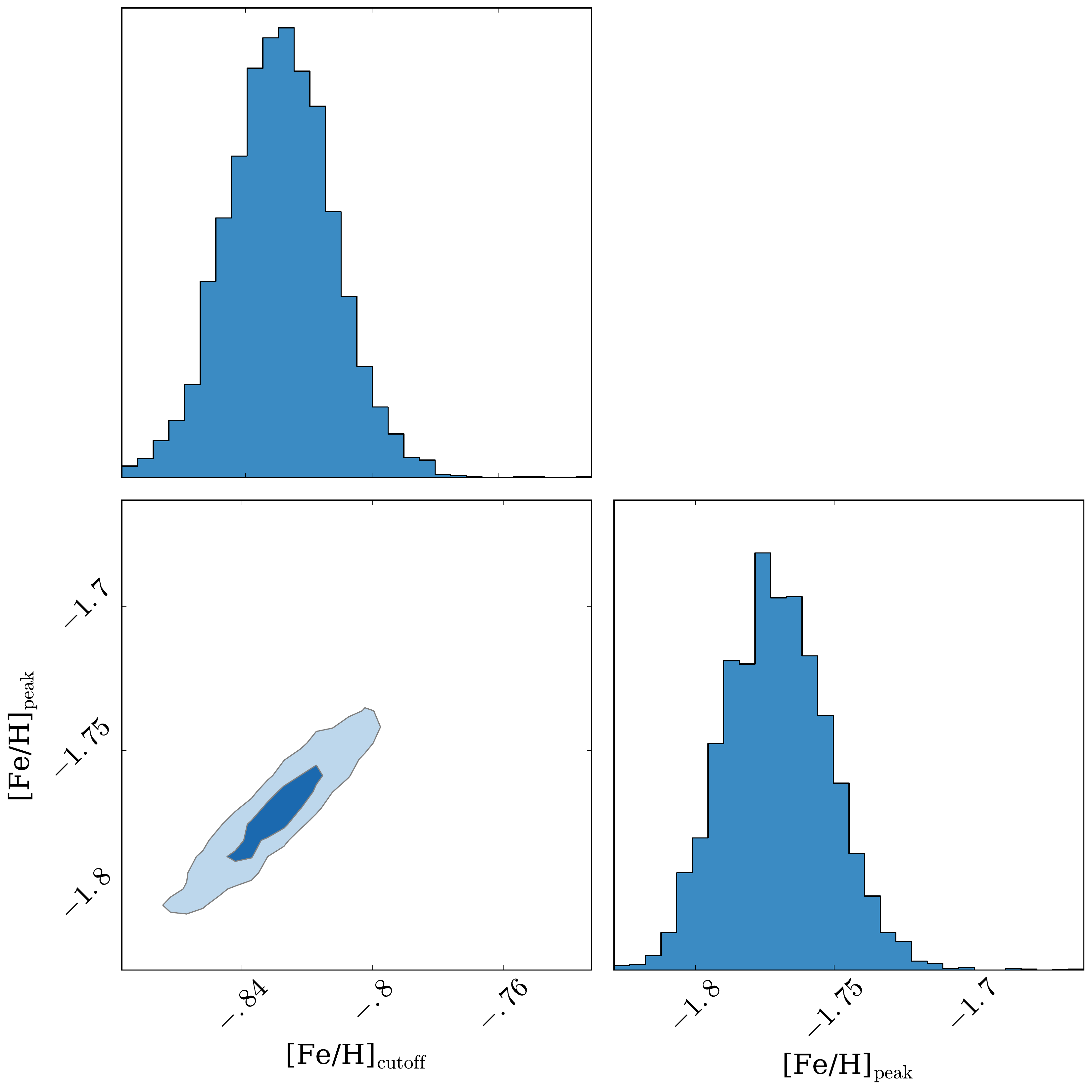}
	\caption{{\tt emcee} results for the metallicity DF of $M_1$, $M_2$, and $M_3$. The plots along the diagonal illustrate 1-D probability distributions for each of the model parameters. The remaining plot shows the joint probability distribution between the two parameters of the metallicity DF. The contours show the 1-$\sigma$ and 2-$\sigma$ confidence levels.\label{fig:emcee_feh}}
\end{figure}
\begin{figure*}
	\centering
	\includegraphics[scale=0.55]{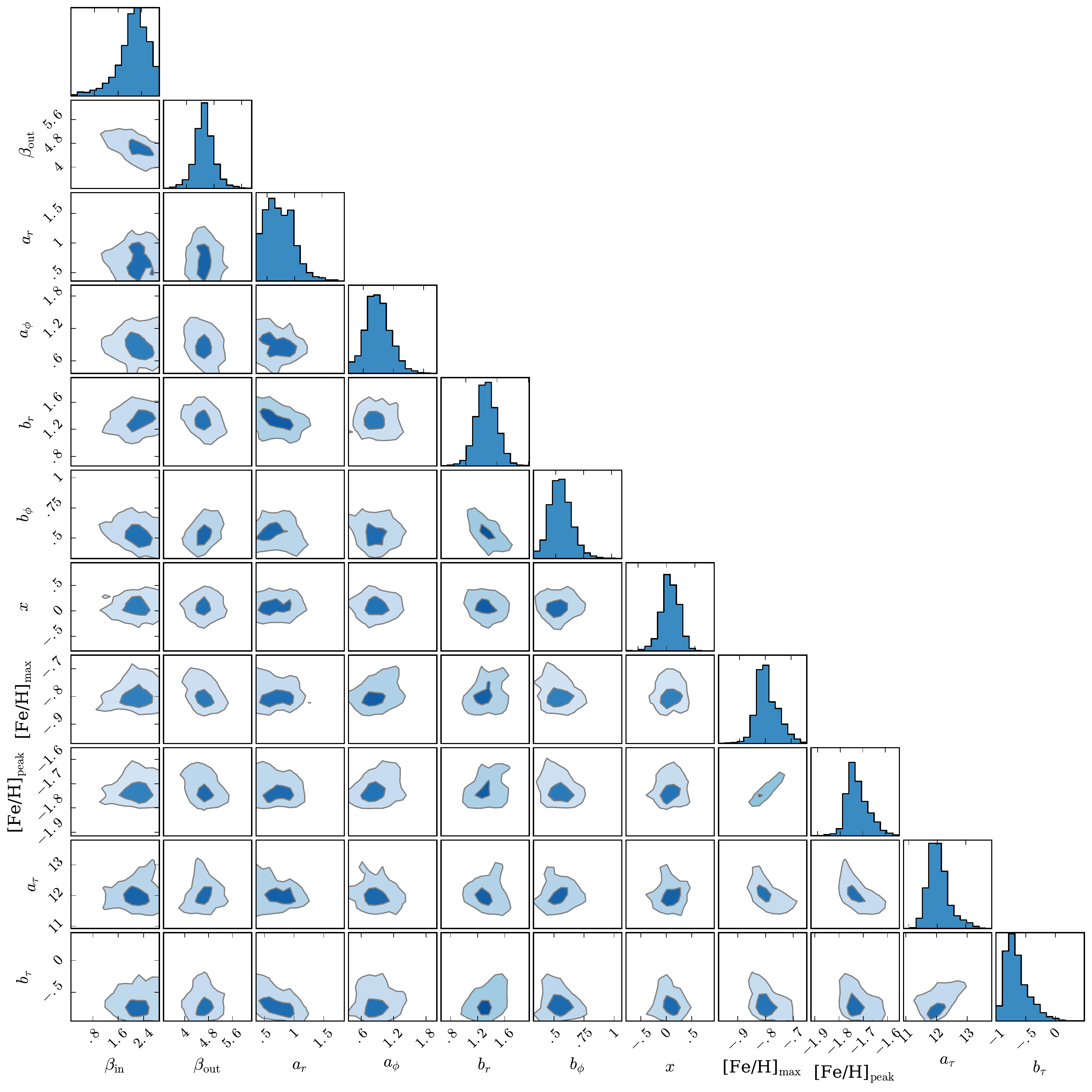}
	\caption{{\tt emcee} results for fitting the EDF $M_4$. The plots along the diagonal are 1-D probability distributions for each of the model parameters. The remaining plots show joint probability distributions between all pairs of parameters. The contours show the 1-$\sigma$ and 2-$\sigma$ confidence levels.\label{fig:emcee_m4}}
\end{figure*}
\subsection[]{Exploring the posterior distribution}
We explore the posterior distribution using two algorithms. The first method is the Nelder-Mead \texttt{amoeba} method \citep{nelder+65}, a downhill-simplex optimization routine for locating an extremum of a multi-dimensional function when the derivatives of that function are not known. This algorithm has been shown to work well in many non-linear optimization problems, although it suffers from the common issue that one may only discover a local maximum, as opposed to a global one (if it exists). The second algorithm we use is the affine-invariant Monte Carlo Markov Chain approach implemented in the \texttt{emcee} package for Python \citep{foreman+13}. The approach uses an interacting ensemble of `walkers' and has been shown to be more effective at dealing with narrow degeneracies than simpler approaches, such as basic Metropolis-Hastings sampling. We now outline our priors and method of
sampling for each of our four models:
\begin{itemize}
\item{\textbf{$M_1$:} Spatial DF and metallicity DF with constant axis ratio (Equation ~\ref{eq:mod1}) and box-uniform distance-metallicity selection function (Equation ~\ref{eqn:chemsfp_const}). We fit all parameters using \texttt{emcee} with 40 walkers and 10,000 steps each. The priors are set as uniform within the following ranges:

\begin{enumerate}
	\item{$0<q<1$}
    \item{$1<\alpha_{\mathrm{in}}<\alpha_{\mathrm{out}}<7$}
    \item{$15<r_{\mathrm{b}}<40$ kpc}
    \item{$0<\FeH_{\mathrm{max}}-\FeH_{\mathrm{peak}}<1$}
\end{enumerate}}
\item{\textbf{$M_2$:} Spatial DF and metallicity DF with variable axis ratio (Equation ~\ref{eq:mod2}) and box-uniform distance-metallicity selection function (Equation ~\ref{eqn:chemsfp_const}). We fit the spatial DF parameters only using \texttt{emcee} with 40 walkers and 10,000 steps each. The metallicity DF parameters are independent and therefore the same as found for $M_1$. The priors are uniform within the following ranges:
\begin{enumerate}
    \item{$q_0>0$}
    \item{$q_{\infty}>0$}
    \item{$0<r_0<50$ kpc}    	
    \item{$0<\alpha<10$}
\end{enumerate}}
\item{\textbf{$M_3$:} Action-based DF and metallicity DF (Equation ~\ref{eq:mod3}) with box-uniform distance-metallicity selection function (Equation ~\ref{eqn:chemsfp_const}). We first fit [$\beta_{\mathrm{in}}$, $\beta_{\mathrm{out}}$, $J_0$] using the \texttt{amoeba} algorithm, with the flattening/anisotropy and rotation parameters fixed at those for an isotropic and round halo, and the metallicity parameters fixed at those found for $M_1$. We then fit [$a_r$,$a_{\phi}$,$b_r$,$b_{\phi}$] using the \texttt{amoeba} algorithm, with the density parameters fixed from the first stage, and metallicity parameters fixed from $M_1$.
The priors are uniform within the following ranges:
\begin{enumerate}
	\item{$0 < \beta_{\mathrm{in}} < 3$}
    \item{$\beta_{\mathrm{out}}>3$}
    \item{$a_{\mathrm{r}}>0.3$}
    \item{$a_{\phi}>0.3$}
    \item{$a_{\mathrm{r}}+a_{\phi}<2.7$}
    \item{$b_{\mathrm{r}}>0.3$}
    \item{$b_{\phi}>0.3$}
    \item{$b_{\mathrm{r}}+b_{\phi}<2.7$}
\end{enumerate}}
\item{\textbf{$M_4$:} Action-based DF, metallicity DF, and age DF (Equation ~\ref{eq:mod4}) with age-varying distance-metallicity selection function. We first fit [$x$, $a_r$, $a_{\phi}$, $b_r$, $b_{\phi}$] using the \texttt{amoeba} algorithm, with the density and metallicity parameters fixed at those found for $M_3$. Then we fit [$a_{\tau}$, $b_{\tau}$] using the \texttt{amoeba} algorithm, with the density, flattening/anisotropy, rotation, and metallicity parameters fixed at those found for the first stage. We then fit all parameters using the \texttt{amoeba} algorithm, and again using \texttt{emcee} with 22 walkers and 500 steps each. We require fewer steps than for $M_1$ and $M_2$  as we start very close to the MAP estimate. However, it cannot be guaranteed that the chains have converged - the limit is simply a function of computational resources (22 walkers and 500 steps take about three weeks to run across 16 cores). The priors are as for $M_2$ with the following set to be uniform in the given ranges:
\begin{enumerate}
	\item{$a_{\tau}<14.5$}
\end{enumerate}}
\end{itemize}
The high-level implementation is in Python and uses the \underline{A}ction-based G\underline{A}laxy \underline{M}odelling \underline{A}rchitecture (\texttt{AGAMA}\footnote{\texttt{AGAMA} can be downloaded from
 \url{https://github.com/GalacticDynamics-Oxford/AGAMA}}). This is a galaxy-modelling library in C++ consisting of several layers that together provide a complete package for constructing galaxy models. The innermost layer provides a range of mathematical tools that include integration, interpolation, multi-dimensional samplers, and units. The central layers provide classes for gravitational potentials, action finders converting phase-space coordinates to actions, and DFs. The outermost layer comprises Python wrappers for several of the C++ classes, and an additional suite of Python routines for fitting DFs, self-consistent modelling, generating mock catalogues, isochrone interpolation, and selection functions. A more detailed description of the library will be given elsewhere (Vasiliev et al. in prep).
\section[]{Results}\label{sec:results}
Here we discuss the favoured parameters determined for our four models, and their uncertainties where derived. We assess the quality of the fit of the models to the observables and investigate the moments associated with $M_4$.
\begin{figure*}
\centering
\includegraphics[scale=0.9]{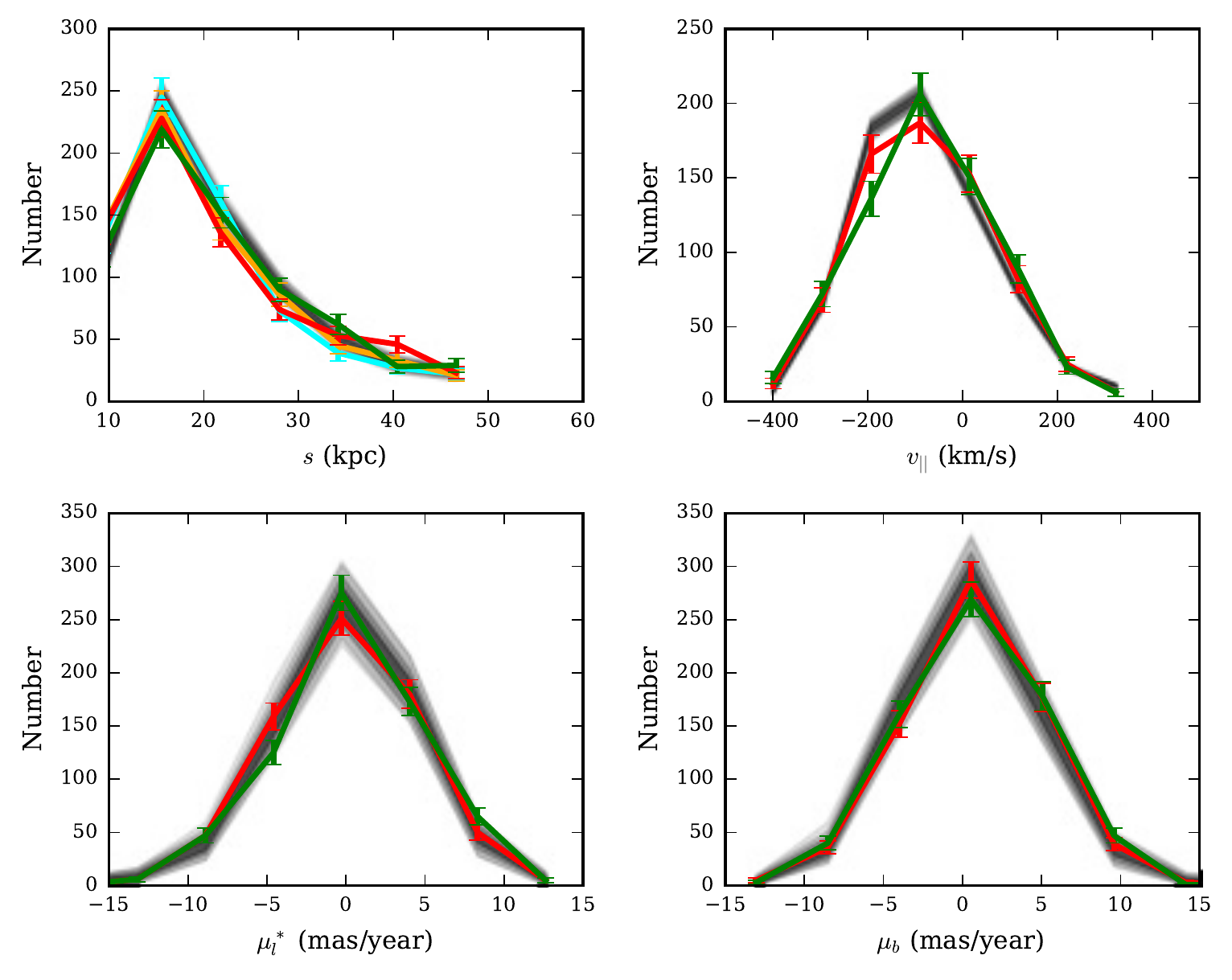}
 \caption{In colour: one-dimensional distributions of mock phase-space
observables (cyan - $M_1$, orange - $M_2$ red - $M_3$, green - $M_4$). The error bars show Poisson errors.
Grey regions: 2000 Monte Carlo resamplings of the data from the error
distributions. Progressively lighter shades of grey indicate $1\sigma$ to $3\sigma$ and 100\% regions for the resampled observables.\label{fig:datafit_1d}}
\end{figure*}
\begin{figure}
\centering
\includegraphics[scale=0.9]{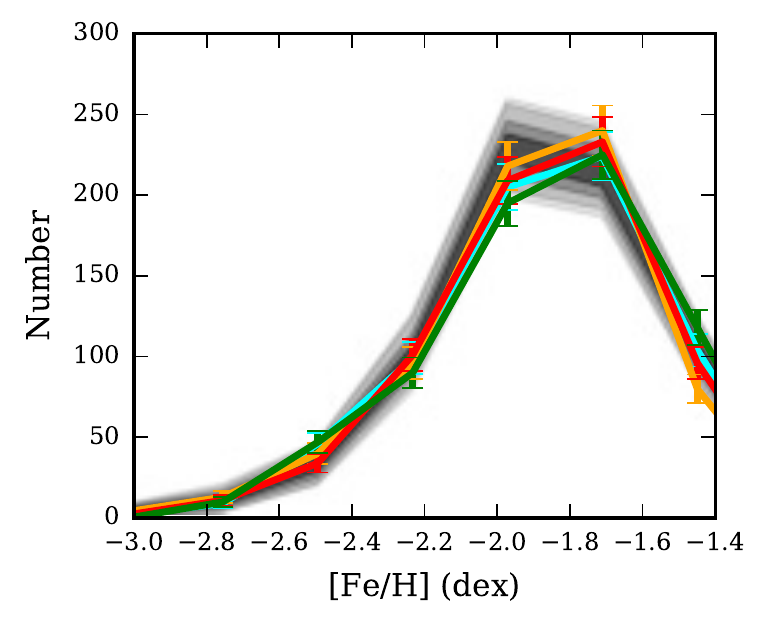}
 \caption{In colour and grey: as Fig.~\ref{fig:datafit_1d} but
 for metallicities.\label{fig:fehfit_1d}}
\end{figure}
\begin{figure*}
\centering
\includegraphics[scale=0.63]{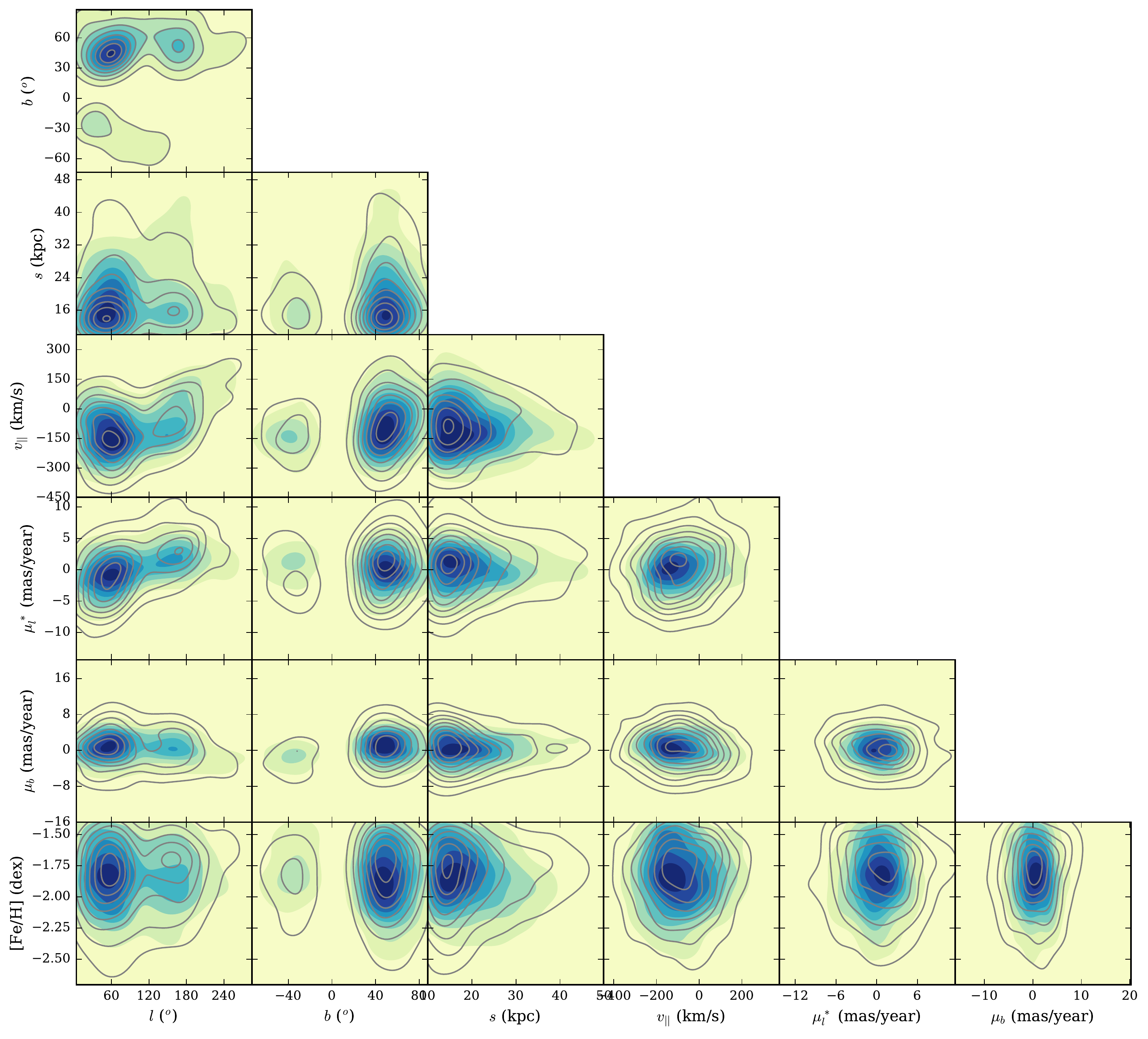}
 \caption{Colour-filled contours: two-dimensional
distributions of measured observables. Black contours: distributions of the mock observables for $\mathrm{M}_4$.
\label{fig:datafit_2d}}
\end{figure*}
\begin{figure}
\centering
\includegraphics[scale=0.55]{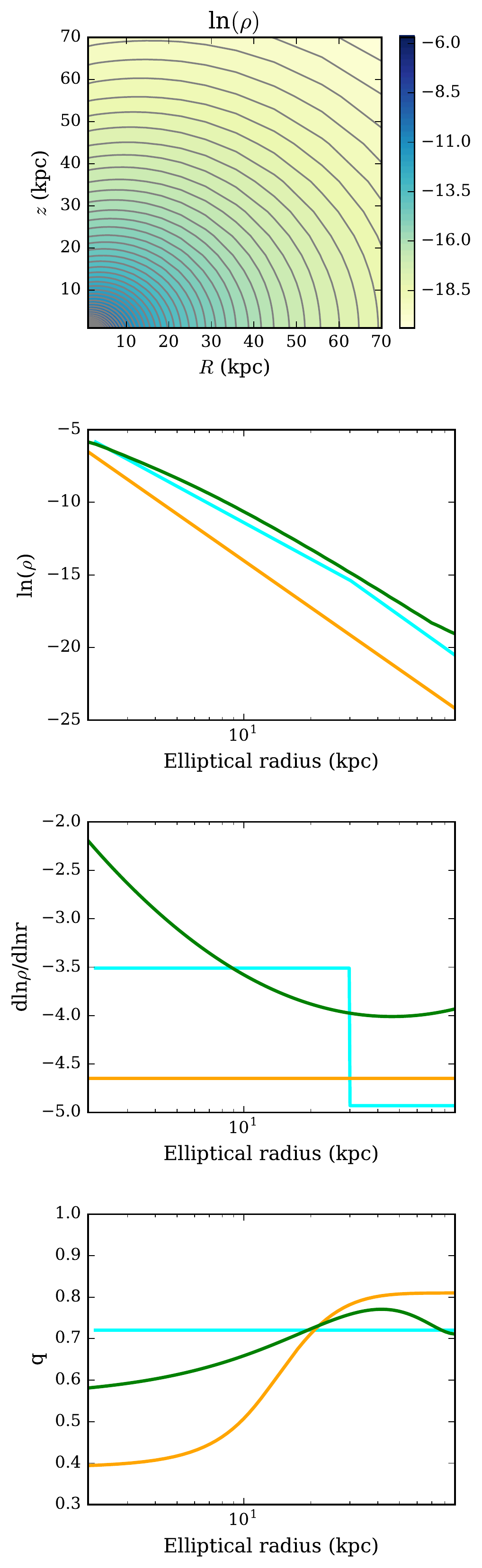}
	\caption{Real-space density: (1) contour map of $M_4$ (top panel), (2) radial density profiles of models $M_1$ (cyan), $M_2$ (orange), and $M_4$ (green) (second panel), (3) logarithmic radial density gradient (third panel), and (4) axis ratio for models $M_1$, $M_2$, and $M_4$ with the same colour coding. \label{fig:model_densitymoment}}
\end{figure}
\subsection[]{The best-fit parameters}
Table~\ref{tab:bestfitpars} gives the median of the recovered parameters for $M_1$, $M_2$, and $M_4$ and the MAP estimates for the parameters after the multi-stage fit for $M_3$. From $M_1$, an axis ratio $q\sim0.7$ is recovered, with a halo that transitions from a power law index of $\alpha_{\mathrm{in}}\sim-3.6$ to $\alpha_{\mathrm{out}}\sim-4.8$ at a break radius $r_{\mathrm{b}}\sim30$ kpc. In the case of $M_2$, the axis ratio varies from $q_0\sim0.4$ to $q_{\infty}\sim0.8$, transitioning at a radius $r_0\sim7$ kpc. The power-law index of the slope is $\alpha\sim-4.7$. Both density profiles would yield a divergent mass if extended towards the centre, and are therefore not physical there. The metallicity DF has a cut-off at $\FeH_{\mathrm{max}} = -0.83$ dex and peaks at $\FeH_{\mathrm{peak}} = -1.77$ dex.

For models $M_3$ and $M_4$, the slope in actions steepens smoothly from $\beta_{\mathrm{in}}\sim-2.1$ to $-2.2$ for actions below $\sim1600$ kpc km s$^{-1}$ to a slope of $\beta_{\mathrm{out}}\sim-4.7$ at larger actions. Allowing a distribution of ages as in $M_4$ results in a minor change in the inner and outer halo power-law indices. Since any rotation within the halo is weak ($x\sim0$), introducing the part in the EDF odd in $J_{\phi}$ probably has little effect on the optimal values of the weights on the actions $a_r, b_r$  and so on. However running an \texttt{emcee} chain, which ensures the parameter space is fully explored, finds different action weights between $M_3$ and $M_4$. In the latter, the isodensity ellipsoids are flattened at low and high actions ($a_{\phi}\ll a_z$ and $b_{\phi}\ll b_z$) and the velocity ellipsoids are elongated in the radial direction in the inner halo ($a_r\ll a_z$). The age model predicts a mean age of $\sim 12.0\,$Gyr, with a negative dependence on actions, i.e. ages decrease outwards. 
\begin{figure*}
\centering
\includegraphics[scale=0.6]{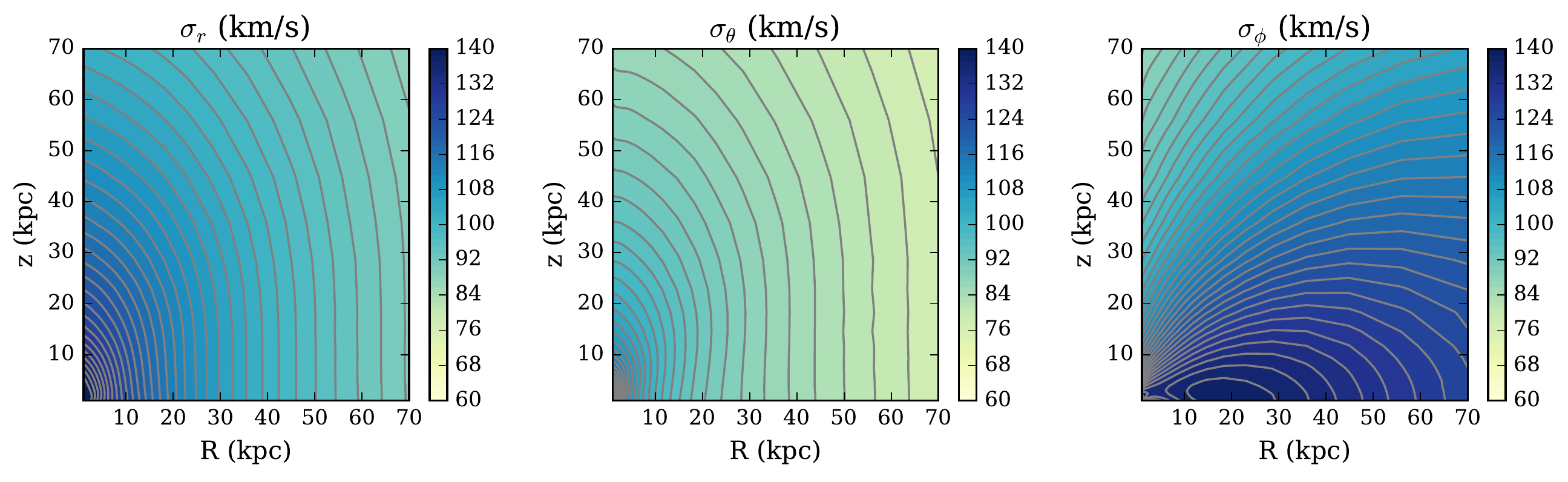}
 \caption{Velocity dispersions predicted by the
 best-fitting EDF. Going from the left to right: spherical radial
velocity dispersion, spherical angular velocity dispersion, and spherical azimuthal velocity dispersion.
\label{fig:model_dispmoments}}
\end{figure*}
\begin{figure}
\centering
\includegraphics[scale=0.7]{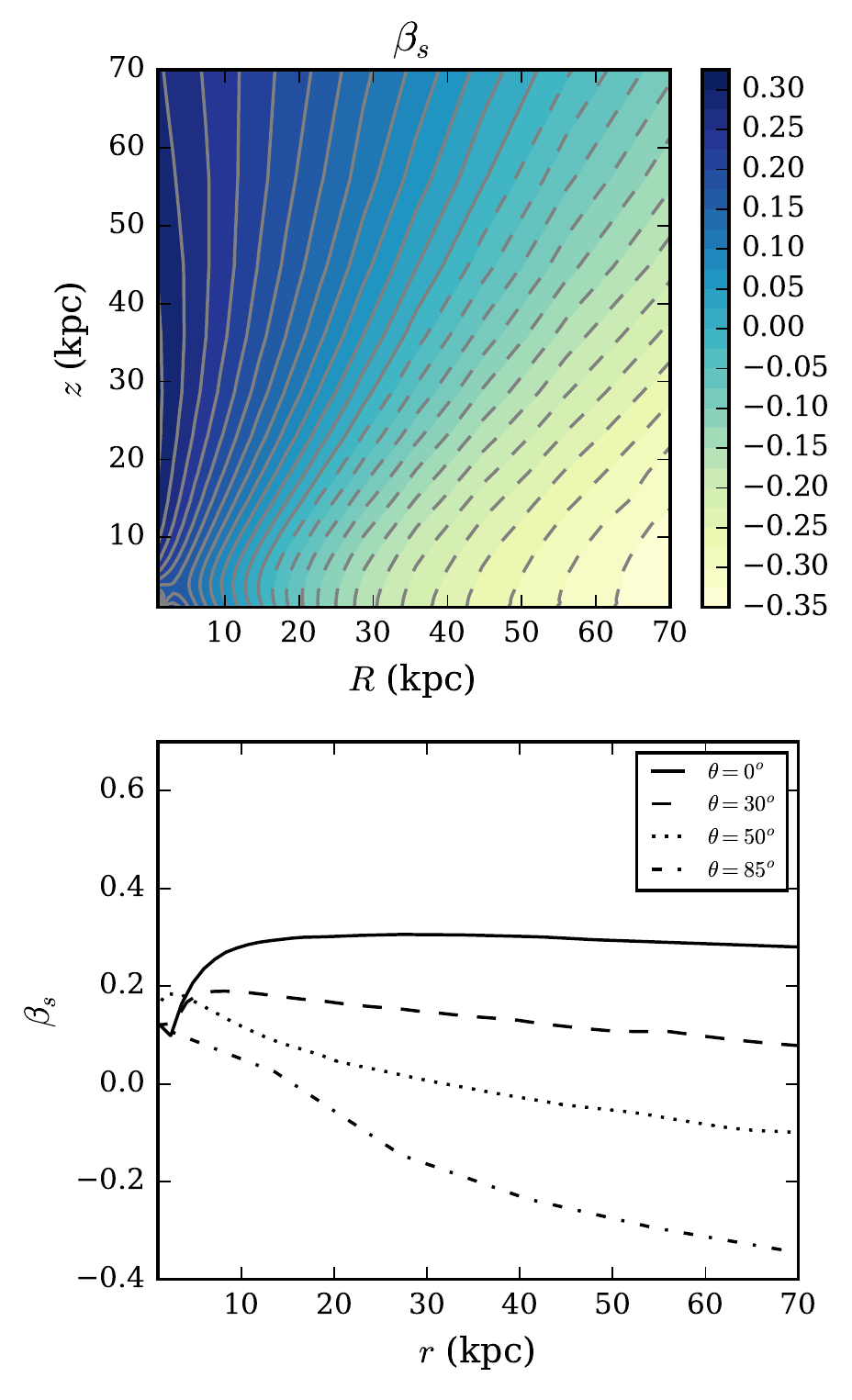}
 \caption{Spherical anisotropy parameter predicted by the best-fitting EDF plotted as a map in $R$ and $z$ (top panel) and against spherical radius for a range of colatitudes $\theta$ (bottom panel).
\label{fig:model_beta}}
\end{figure}
\subsection[]{Uncertainties on the recovered parameters} 
Table~\ref{tab:bestfitpars} gives the 68\% confidence intervals for parameters of $M_1$, $M_2$, $M_3$ (metallicity DF parameters only) and $M_4$. Figs.~\ref{fig:emcee_m1}, ~\ref{fig:emcee_m2}, ~\ref{fig:emcee_feh}, and ~\ref{fig:emcee_m4} present the 1-D and joint probability distributions from the {\tt emcee} runs. The metallicity DF is independent of the spatial DF in $M_1$, $M_2$ and of the phase-space DF in $M_3$. Its parameters are thus shown separately.

The uncertainties on the parameters of $M_1$ and $M_2$ are of the order of $\sim5$--$10\%$ for the spatial DF, except for the inner axis ratio and flattening transition radius of $M_2$, which are more uncertain ($\sim 20\%$). There are positive correlations between the outer slope and between the break radius of $M_1$, and the outer axis ratio, flattening transition radius, and slope of $M_2$.

For models $M_1$, $M_2$, and $M_3$, the uncertainties on the maximum and peak metallicities are smaller, at the level of $\sim1$--$2\%$, showing that they are well constrained. Fig.~\ref{fig:emcee_feh} shows a correlation between the two metallicity parameters that arises from the difference between them being limited to unity.

The uncertainties on the parameters of $M_4$ vary greatly. Those on the parameters of the metallicity DF are greater than the uncertainties in $M_1$ and $M_2$ because of the flexibility introduced with the age model, which modifies the distance-metallicity selection function. The uncertainties on the action weights vary between $\sim 15$ to $35 \%$. The uncertainty on the rotation parameter is high and translates to a 68\% confidence interval $\sim[-10, 30]$ kms$^{-1}$ for the rotation speed.

The mean age in the case of no dependence on actions has an uncertainty $\sim0.33\,$Gyr, primarily towards higher ages, because the distance-metallicity selection function varies much less there. The dependence on age is quite uncertain, again partly because of the insensitivity of the distance-metallicity selection function to the oldest ages. 

The 1-D marginalized probability distributions in Fig.~\ref{fig:emcee_m4} are unimodal, but often skewed. Asymmetrical distributions include the mean age ($a_{\tau}$) in the case of no dependence on actions, and the dependence on actions ($b_{\tau}$), both of which have an extended tail towards higher values. The inner halo power-law index ($\beta_{\mathrm{in}}$) has an asymmetrical distribution with an extended tail towards lower values. Several correlations exist between parameters. There is a weak, negative correlation between the power-law indices of the inner and outer halo ($\beta_{\mathrm{in}}$ and $\beta_{\mathrm{out}}$), i.e. a lower value of the index in the inner halo can be partly compensated by a larger value of the index in the outer halo. There is also a correlation between the weight on the radial action in the outer halo ($b_r$) and the weight on the angular action there ($b_{\phi}$), and to a lesser extent in the inner halo. This highlights a connection between the flattening and anisotropy, i.e. more flattened systems tend to have a higher degree of radial anisotropy. There is also the correlation between the parameters of the metallicity DF. Other correlations may also exist but are obscured by the lack of sufficient resolution in the \texttt{emcee} runs.  
\subsection[]{Fits to the observables}\label{ssec:fits}
We assess the quality of the model fits by generating mock catalogues at the measured sky positions, using an adaptive sampling-rejection method from \texttt{AGAMA}. The product of the SF and EDF is sampled in $\log (s/\mathrm{kpc}) $, $\log \FeH$, and for $M_3$ and $M_4$ in $\tanh (v_r/\mathrm{km\,s}^{-1})$, $\tanh(\mu_l^*/\mathrm{mas})$, and $\tanh(\mu_b/\mathrm{mas})$. In these coordinates the value of the EDF varies less strongly. Each proposed sample has noise added to it according to the errors measured on the observables at those sky positions. The parameters used to generate the mock samples in each case correspond to the MAP estimates.

The coloured points (cyan - $M_1$, orange - $M_2$, red - $M_3$, and green - $M_4$) joined by lines in Figs.~\ref{fig:datafit_1d}
and \ref{fig:fehfit_1d} show histograms of the mock catalogues. The region
covered by analogous histograms of 2000 resamplings of the error
distributions of the measured observables are shown by the grey regions. In plots for observables with small errors, such as $l$, $b$, $s$, and $v_\parallel$, the grey regions form fairly well defined curves. In plots for observables with large errors, namely $(\mu_l^*,\mu_b,\FeH)$, the grey regions fill out a region of significant width. The coloured curves generally overlap with this region, indicating that the EDF and SF are together
doing a good job at describing the location of the observables. 

Fig.~\ref{fig:datafit_2d} compares histograms for the joint distribution of
pairs of phase-space observables $(l,b)$, $(s,b)$, $(\FeH,b)$, etc. in the case of parameters corresponding to the MAP estimate obtained for $M_4$. The colour-filled contours show
distributions of mock observables, while the black contours show the
distributions of measured observables. In general there is good agreement
between the mock and observational distributions.
\subsection[]{Moments of the recovered parameters of $M_4$}\label{ssec:moments}
We now describe the model $M_4$ with the parameters corresponding to the MAP estimate.
\subsubsection[]{Density of stars in real space}
Fig.~\ref{fig:model_densitymoment} shows the shape of the density distribution. The colour scale in the top panel shows $\rho(R,z)$. Flattening of the contours is evident. By fitting ellipses to the isodensity curves, we obtain the radial density profile shown by the green curve in the second panel. The steepening of the density profile with increasing radius is shown by the green curve in the third panel, which gives the logarithmic radial density gradient. The slope steepens smoothly from $-2.2$ at $\sim 2$ kpc to $\sim -4$ in the outer halo. The green curve in the bottom panel shows that the halo is flattened ($q\simeq0.6$ to $0.8$) throughout. Radial profiles of the logarithmic density, logarithmic density gradient, and axis ratio are shown also for $M_1$ (cyan) and $M_2$ (orange), each plotted against elliptical radius. The density profile of $M_1$ is steeper in the inner and outermost parts than that of $M_4$, while that of $M_2$ is steeper throughout. The axis ratio of $M_1$ is very similar to that of $M_4$. The axis ratio of $M_2$ is significantly lower than that of $M_1$ and $M_4$ in the inner halo and moderately higher in the outer halo.
\subsubsection[]{The velocity ellipsoid}

Fig.~\ref{fig:model_dispmoments} shows the velocity dispersions of $M_4$. $\sigma_r$ generally dominates at high $z$ throughout, implying radial anisotropy there. At low $z$, $\sigma_{\phi}$ dominates, implying tangential anisotropy. The contours of constant $\sigma_r$ and $\sigma_{\theta}$ are elongated in the $z$ direction, while $\sigma_{\phi}$ contours are elongated in the $R$ direction. Fig.~\ref{fig:model_beta} shows the spherical anisotropy parameter
\begin{equation}
\beta_{\rm s}= 1 - \frac{\sigma_{\theta}^2 + \sigma_{\phi}^2}{2\sigma_r^2}.
\end{equation}
The degree of radial anisotropy increases from a tangential bias in the equatorial plane to $\sim0.3$ at the highest point along the $z$-axis. The lower panel of Fig.~\ref{fig:model_beta} shows radial profiles of the spherical anisotropy parameter against spherical radius along a range of polar angles, $\theta$, where $\theta = 0$ is along the $z$ axis and $\theta = 90^\mathrm{o}$ is in the equatorial plane. In the equatorial plane, the orbits vary from isotropic to mildly tangential. Nearer the $z$ axis, the profiles become more radially anisotropic.

\subsubsection[]{The distribution of ages}
\begin{figure}
\centering
\includegraphics[scale=0.6]{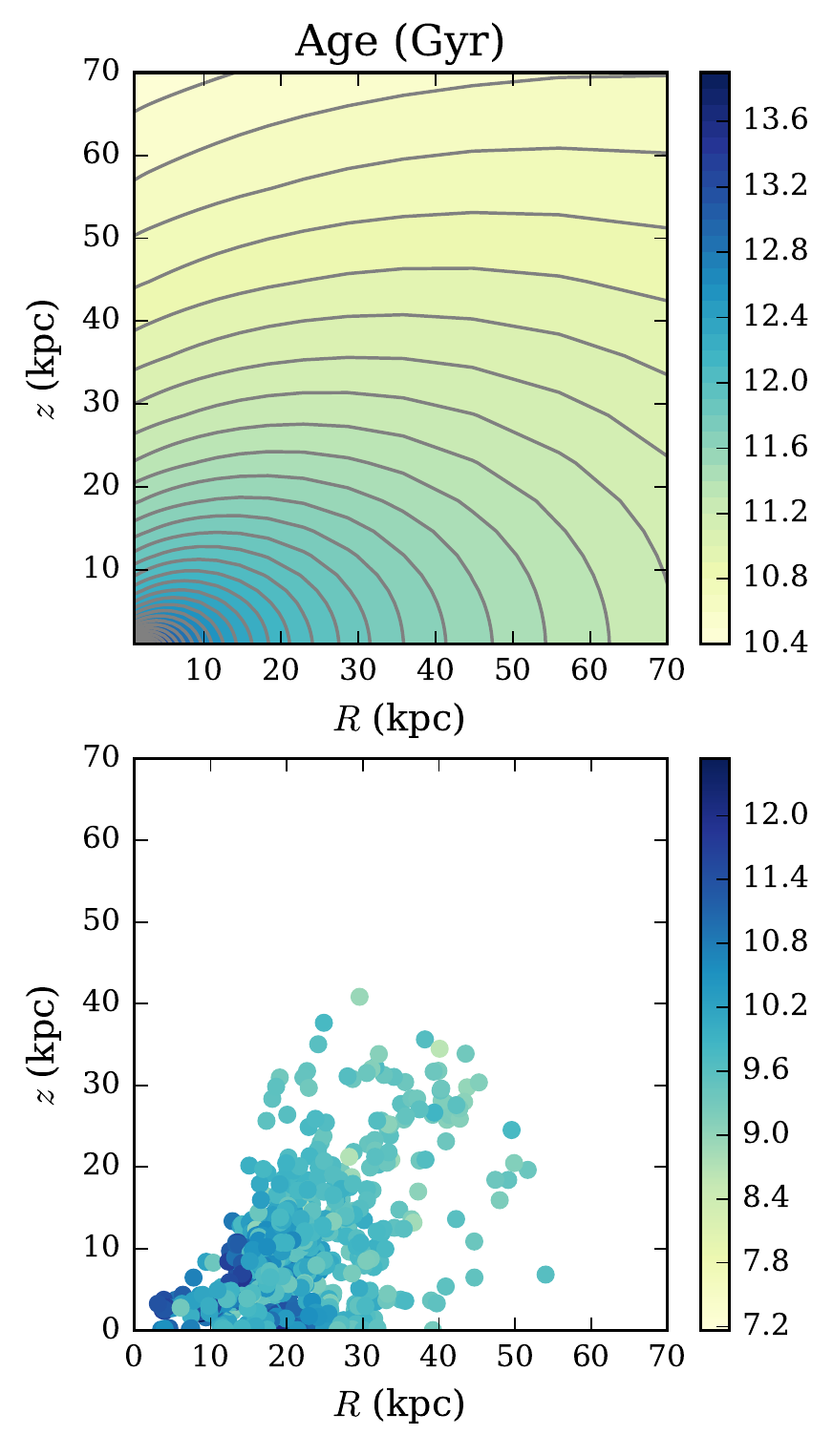}
 \caption{Age map predicted from the total action moment (top panel) and for the stars in the SEGUE-II sample (bottom panel).
\label{fig:model_age}}
\end{figure}

The top panel of Fig.~\ref{fig:model_age} shows the distribution of mean ages. Contours of constant mean age are flattened, approximately as the contours of constant density. The bottom panel shows the age map inferred from the actions of the stars in the SEGUE-II sample, convolved with their error distributions. In general stars at lower $R$ and $z$ have higher ages, but a high velocity implies large actions and therefore a relatively young age. Therefore the gradient in age with actions manifests as a small negative age gradient with radius $\sim-0.03\,$Gyr kpc$^{-1}$. The plot also suggests that we are biased towards observing younger stars.
\section[]{Discussion}\label{sec:discuss}
Here we focus our discussion on the results of fitting the full phase-space EDFs, how they compare to the literature, and highlight uncertainties that may impact these results.
\subsection[]{Our perspective on the stellar halo}
\subsubsection[]{Distribution of stars in action space}
Traditionally metallicity and age gradients have been examined as a function of radius. However, stars are better characterised by their actions, which do not vary along orbits. A clear separation in action space becomes `smeared' in real space, and a glance at the picture in action space can be enlightening. We find that the EDF declines more rapidly with actions in the outer halo (slope $\sim-5$) than in the inner halo (slope $\sim-2$). The weights on the actions in the EDF suggest a flattened stellar system, which is tangential to isotropic in the equatorial plane, becoming more radially anisotropic as $z$ increases. The part of the EDF odd in $J_{\phi}$ is negligible. We find the ages of the stars to be well predicted by a log-linear dependence on the total action, with higher ages at smaller actions. The gradient is thus negative ($-0.69\,\mathrm{Gyr}\,\mathrm{dex}^{-1}$). A single-age model is, however, also able to reproduce the observations.

\subsubsection[]{Distribution of stars in real space}
%
The slopes in action space translate to a density profile in real space that steepens with radius from a slope of $\sim-2$ at $\sim 2$ kpc  to $\sim-4$ by 30 kpc. The gradient in ages with actions translates to a real-space gradient $\sim -0.03\,$Gyr kpc$^{-1}$, subject to a significant degree of uncertainty.

The weights on actions determine the shape of the density and velocity ellipsoids. The halo's axis ratio is roughly constant at $\sim0.7$, very similar to what is implied by the broken power-law model. The orbital structure varies from mildly tangential to moderately radially anisotropic throughout, becoming more radial as you move from the equatorial plane towards the $z$ axis.

The negligible part of the EDF that is odd in $J_{\phi}$ generates a very small level of rotation with a large uncertainty (our 68\% confidence interval extends between $-10$ to $30$ km s$^{-1}$).

Finally, we did not probe the distribution of metallicities in action space directly. It is however linked to the EDF through the selection function in distance and metallicity. The metallicity DF is well described by a single lognormal distribution with a peak at $\sim-1.8$ dex and with a maximum at $\sim-0.8$ dex.
\subsubsection[]{Is there a difference between the inner and outer halo?}
There is compelling evidence for differences between the inner and outer halo in action space, primarily due to the difference in the inner and outer slopes in actions. This manifests in real space as a steepening of the density profile with radius, a non-negligible negative age gradient with radius, and a variation in the anisotropy with radius.

There have been several claims of two populations in the halo \citep[e.g.][]{carollo+07,beers+12,deason+13,hattori+13}. Although our sample is too small to rule out such a dichotomy, our models, which predict smooth transitions between the inner and outer halo, are sufficient to reproduce the current data.
\subsection[]{Comparison with previous work}
\subsubsection[]{Radial density profile}
Several authors have determined density profiles for halo BHBs, finding a power-law index of $\sim-2.5$ to$-3.5$ \citep[e.g.][]{preston+91,kinman+94,sluis+98,dePropis+10}. Double power-law profiles have also been fitted to BHBs \citep{deason+11a}, finding a power-law index of $-2.3$ in the inner halo, $-4.6$ in the outer halo, and a break radius of $27$ kpc. The first two power-law indices are very similar to our findings. Similar density profiles have been recovered for other stellar populations probing deep into the halo, such as RR Lyrae \citep{watkins+09,sesar+13}, subdwarfs \citep{smith+09b} and K giants \citep{xue+15,das+16}. An axis ratio ranging from $0.5$--$0.6$ has been found in BHBs \citep[e.g.][]{sluis+98,deason+11a} and $0.6$--$1.0$ in other stellar populations \citep[e.g.][]{carollo+07,watkins+09,sesar+13,xue+15,smith+09b}.
\subsubsection[]{The velocity ellipsoid}
\cite{deason+11b} and \cite{hattori+13} found the BHB stars in the Milky Way halo to exhibit a dichotomy between a prograde-rotating, comparatively metal-rich component ($\FeH>−2$) and a retrograde-rotating, comparatively metal-poor ($\FeH<−2$) component. \cite{deason+11b} attribute the prograde metal-rich population to the accretion of a massive satellite ($\sim10^9$M$_{\odot}$) and the metal-poor population to the primordial stellar halo. The net retrograde rotation might then reflect an underestimate in the adopted LSR circular velocity. \cite{fermani+13b} however  remeasured the rotation of the Milky Way stellar halo on two samples of BHB halo stars from SDSS with four different methods, and found a weakly prograde or non-rotating halo in all cases. They attributed the rotation gradient across metallicity measured by \cite{deason+11b} on a similar sample of BHB stars to the inclusion of regions in the apparent magnitude-surface gravity plane known to be contaminated by substructures. \cite{sirko+04} did not find any rotation in their sample of BHBs and \cite{smith+09b} do not find any rotation in their sample of subdwarfs.

Several authors \citep{deason+12a,kafle+12,williams+15b,cunningham+16} found a radially biased velocity ellipsoid overall but some find a region around $\sim20$ kpc with tangential anisotropy. From a similar sample of BHBs, \cite{sirko+04} found isotropy and \cite{hattori+13} found the metal-rich component to exhibit mild radial anisotropy, and the metal-poor component to exhibit tangential anisotropy. Analysis of other halo tracer populations have also arrived at a mixture of conclusions. \cite{dehnen+06} found radial anisotropy in a mixture of globular clusters, horizontal-branch and red-giant stars, and dwarf spheroidal satellites. \cite{smith+09a} found the velocity ellipsoid of SDSS subdwarfs to be radially biased. \cite{carollo+10} found inner-halo metal-richer stars on radially anisotropic orbits, and outer-halo stars to be on less eccentric orbits. 

The diversity in conclusions regarding halo anisotropy may be a result of several things. Spectroscopic surveys can have a strong selection bias on metallicity (not so much in BHBs, but in K giants, see \cite{das+16}), and if there is a correlation between metallicity and dynamics, then a lack of treatment of such a bias can lead to erroneous conclusions about anisotropy. It is also true that the proper motions currently available are highly uncertain. Furthermore, it should be emphasised that our models are designed to fit the smooth, phase-mixed halo, and this may be why we do not reproduce the `dip' in anisotropy found by several authors at $20$ kpc. On a related note, we have attempted to remove substructures where possible - the exact samples used by various authors differ slightly in their observed velocity distributions and thus derive different anisotropy profiles.

Our model qualitatively agrees with simulations at high $z$ (i.e. radial bias  increases outward), though with a lesser degree of radial bias throughout. In such simulated haloes, the primary mechanism for growth since $z\sim2$ is thought to be accretion onto the halo through minor mergers \citep{bullock+05,abadi+06}, and accreting objects have rather radial orbits.
\subsubsection[]{The distribution of chemical properties}
\cite{carollo+07} and \cite{beers+12} found evidence for two metallicity components in a mixed sample of stellar types. \cite{an+15} analysed a sample of main-sequence halo stars and found two components peaking at [Fe/H]$\sim-1.7$ and $\sim-2.3$. \cite{xue+15} and \cite{das+16} reached similar conclusions about SDSS K giants.

We find one lognormal component is sufficient to describe the metallicity DF of the BHBs. The discrepancy may arise from a difference between the types of systems thought to contribute to halo stars. There is an age-metallicity bimodality in the Milky Way globular cluster system \cite[e.g.][]{leaman+13}. \cite{fiorentino+15} analysed the periods and luminosity amplitudes of field RR Lyrae stars and found that dwarf spheroidals lacked high-amplitude short-period variable stars; whereas these are found in globular clusters and massive dwarf irregulars such as the Sagittarius stream. 

 \cite{preston+91} detected a colour gradient in BHBs out to $\sim12$ kpc, which \cite{santucci+15} consolidated with a larger sample, extending out to $\sim40$ kpc. They claimed that the 
gradient is independent of metallicity and therefore indicate a gradient in age. More massive systems penetrate deeper into the gravitational potential. We also expect these systems to have the oldest components because they would have grown over a longer timescale. From a similar argument however, we would also expect chemical evolution to have occurred more rapidly in these systems, therefore producing a metallicity gradient. It is unclear why the difference in the make-up of the inner and outer halos would manifest solely as an age gradient rather than a metallicity gradient. The metallicity gradient could just be small.
\subsection[]{Uncertainties in the analysis}
The errors quoted in this work represent statistical uncertainties only, rather than systematic errors, which are more difficult to characterise. We discuss possible sources of systematic errors below.
\subsubsection[]{Impact of resonances and chaos}
We have assumed a fully-integrable potential in which the number of isolating integrals of motion is equal to the number of degrees of freedom. In such cases a transformation from phase-space coordinates to angle-action coordinates can be done globally. Real potentials however permit families of resonantly trapped orbits. These resonant orbits possess actions, but require a different transformation. In non-integrable potentials, some fraction of the orbits will be chaotic. Chaotic orbits are not bound to the surface of a torus and instead fill the spaces between tori. Chaotic orbits have no orbital actions. \cite{binney16} concluded that the net impact of resonant trapping on the dynamics of halo stars is likely to be small. 
\subsubsection[]{A fixed potential and parametrised EDF}
An EDF can perfectly model the data only in the true potential. Therefore any imperfections in our choice of potential will both bias our EDF away from the true EDF and give rise to discrepancies between our best model and the data. Our success in reproducing the data suggests that our chosen potential is not seriously in error.  
The supposition of a particular functional form for the EDF can bias the results by restricting the set of possible solutions, despite allowing a range of density and anisotropy profiles. Our ansatz regarding the dependence of stellar ages on actions represents just one, physically-motivated, possibility that is simple to calculate. An age gradient is not forced by the EDF however; if none were needed by the data, age would have been found to be independent of actions.
\subsubsection[]{Stellar population assumptions}
Our evaluation of the distance-metallicity selection function depended on  relations from isochrones between the age, mass, and metallicity of a star and its luminosity in various wavebands. Systematic errors arising from faulty isochrones are difficult to assess. The ability of our EDF to produce a similar density profile for the BHBs to that in the literature suggests that our selection function is not significantly in error.
\subsubsection[]{Impact of substructure}
We have masked the Sagittarius stream in this analysis, but we do not know how other, unmasked substructures may impact our assessment of the halo's structure. Our ability to sufficiently reproduce the phase-space observables after excluding the stream implies that either we are predominantly probing the smooth stellar halo with the data or the current data are too sparse to resolve the halo's substructure. 
\section[]{Conclusions and further work}\label{sec:conc}
We probed the chemodynamical structure of Milky Way halo BHBs by combining spatial and action-based EDFs that describe the locations of stars in phase space, metallicity, and age. The analysis allows a more natural description of the ages of BHBs in action space in which their separation is clearer than in real and velocity space. The specification of an EDF enables the incorporation of a realistic selection function that takes into account restrictions on sky positions, apparent magnitudes, and colours. In general, our models reproduce the observations well. This may be an argument that there is enough phase-mixed debris for action space to be smoothly populated (at least for relatively tightly bound orbits). i.e., there may be a part of the inner halo that will always be well represented by smooth models. Alternatively it may be because the data for BHBs are not yet rich enough to resolve most halo substructures. 

The EDF of the BHBs is steeper at larger actions than at smaller actions. Older stars are found at smaller actions and younger stars at larger actions. The spatial distribution of the stars is similarly well reproduced by a broken power law with a constant axis ratio, a single power law with a variable axis ratio, and a gradually steepening power law with a variable axis ratio. Fitting positions and velocities simultaneously yields a density profile that steepens smoothly from $\sim-2$ at $\sim 2$ kpc to $\sim-4$ in the outer halo. The halo is moderately flattened with an axis ratio $\sim 0.7$ throughout. The overall metallicity distribution is well described by a single lognormal component that has a maximum metallicity at $\sim-0.8$ dex and a peak at $\sim-1.8$ dex. Our full phase-space EDF also allowed rotation - this could be at a level of $-10\,$km s$^{-1}$ to $30\,$km s$^{-1}$ at most but the median result favours no rotation. The stellar velocity ellipsoid varies from tangential bias in the equatorial plane to radially elongated at high $z$. Allowing a dependence of stellar ages on actions leads to an age gradient $\sim-0.03$ kpc$^{-1}$, with moderate uncertainty. However, an EDF assuming approximately a single age of $11\,$Gyr is also able to fit the observables well.

There are several possible directions for further work. The EDF could be applied to detect substructures in a richer sample of halo stars in the phase-space-metallicity domain. The EDF could be changed to make the transition between the inner and outer asymptotic slopes of the density profile sharper. The EDF could be further elaborated to include a dependence on [$\alpha$/Fe]. 

\section*{Acknowledgements}
The research leading to these results has received funding from the European Research Council under the European Union's Seventh Framework Programme (FP7/2007-2013)/ERC grant agreement no.\ 321067. PD thanks GitHub for providing free private repositories for educational use. AW acknowledges the support of STFC. PD is also grateful for fruitful discussions with members of the Oxford Galactic Dynamics group.

\bibliographystyle{mnras}
\bibliography{biblio}

\appendix

\label{lastpage}

\end{document}